\documentclass[prb,twocolumn,showpacs,preprintnumbers,amsmath,amssymb,floats]{revtex4-1}
\usepackage{graphicx}
\usepackage{amsmath}
\usepackage{epstopdf}
\usepackage{amsbsy}
\usepackage{color}
\usepackage{lipsum}
\usepackage{dcolumn}
\usepackage{bm}

\newcommand{\eq}[2]{\begin{equation}\begin{aligned}\label{#2}#1\end{aligned}\end{equation}}
\renewcommand{\eqref}[1]{Eq.\ (\ref{#1})}
\renewcommand{\k}{{\mathbf{k}}}
\newcommand{\ev}[1]{\langle #1 \rangle}
\newcommand{\evt}[1]{\langle T_\tau #1 \rangle}
\renewcommand{\oc}[1]{c_{{#1}}} 
\newcommand{\ocd}[1]{c_{{#1}}^{\dagger}}  
\newcommand{\sgn}{\operatorname{sgn}}

\usepackage{tikz}
\usetikzlibrary{calc,decorations.markings}
\usetikzlibrary{snakes}
\newcommand{\vertex}[1]{\shade[ball color=black] #1 circle (0.12);}

\begin{document}
\title{Spin excitations in the nematic phase and the metallic stripe spin-density wave phase of iron pnictides}
\author{M. Kovacic, M. H. Christensen, M. N. Gastiasoro, and B. M. Andersen}
\affiliation{Niels Bohr Institute, University of Copenhagen, Universitetsparken 5, DK-2100 Copenhagen,
Denmark}
\date{\today}
\begin{abstract}
We present a general study of the magnetic excitations within a weak-coupling five-orbital model relevant to itinerant iron pnictides. As a function of enhanced electronic correlations, the spin excitations in the symmetry broken spin-density wave phase evolve from broad low-energy modes in the limit of weak interactions to sharply dispersing spin wave prevailing to higher energies at larger interaction strengths. We show how the resulting spin response at high energies depends qualitatively on the magnitude of the interactions. We also calculate the magnetic excitations in the nematic phase by including an orbital splitting, and find a pronounced $C_2$ symmetric excitation spectrum right above the transition to long-range magnetic order. Finally, we discuss the $C_2$ versus $C_4$ symmetry of the spin excitations as a function of energy for both the nematic and the spin-density wave phase.
\end{abstract}


\maketitle

\section{introduction}

The proximity of magnetism appears important for the existence of unconventional superconductivity, suggesting that magnetic fluctuations may play a crucial role in stabilizing the superconducting state. This scenario has naturally highlighted the importance of detailed experimental and theoretical studies of spin fluctuations in heavy fermion materials and high-temperature superconductors.\cite{scalapino12} In these systems, prominent short-range magnetic fluctuations remain in the normal state obtained by destruction of the magnetic order by doping, and a magnetic inelastic resonance mode emerges in the superconducting state. For the cuprates, worldwide neutron scattering studies have led to the discovery of the so-called hour-glass dispersion and its associated doping and temperature dependence.\cite{cupratereview2}

Recently much attention has focussed on the magnetic properties of iron pnictides and iron chalcogenides.\cite{lumsden10} In the iron pnictides, the magnetic structure is of the so- called $\mathbf{Q}_1 = (\pi,0)$ collinear stripe order, consisting of in-plane moments oriented antiferromagnetically (ferromagnetically) along the $a$ ($b$) axis of the orthorhombic 1-Fe lattice. Other related magnetic structures including double-Q order consisting of superpositions of $\mathbf{Q}_1$ and $\mathbf{Q}_2 = (0,\pi)$ are, however, close by in energy and may be realized in some of these materials.\cite{lorenzana08,eremin2010,kim10,brydon11,inosov13,avci14,gastiasoroMn,gastiasoroC4} The potential existence of the double-Q phases is important to settle since it provides indirect evidence for a phase diagram determined mainly by itinerant magnetic interactions. It was recently proposed to use magnetic excitations as a means to detect the fingerprints of the double-Q phases.\cite{wang14}

The microscopic nature of magnetism in iron-based superconductors has been extensively discussed in the context of the small structural tetragonal-to-orthorhombic lattice distortion at $T_S$ which tracks but pre-empts the magnetic N\'{e}el transition at $T_N < T_S$ in many of the iron pnictides. In the itinerant magnetic picture, the structural transition is caused by a magneto-elastic coupling and a magnetic Ising-nematic transition at $T_S$ breaking only the $Z_2$ symmetry of the $Z_2 \times SO(3)$ symmetric paramagnetic phase.\cite{fang08,xu08,hu12,fernandes12} By contrast, in the orbital ordering scenario, orbital order sets in at $T_S$ which modifies the exchange couplings between the Fe ions and thereby trigger a transition to the $\mathbf{Q}_1$ stripe magnetic phase at lower temperatures.\cite{kruger09,chen09,lv09,lee09}

Experimentally, the magnetic excitation spectrum of the iron pnictides has been studied extensively throughout the phase diagram with a prominent neutron resonance mode dominating the response in the pure superconducting phase at $T < T_c$.\cite{lumsden10} In the nematic paramagnetic phase ($T_N < T < T_S$) several recent experiments have accessed the spin excitation anisotropy.\cite{harriger11,fu12,luo13,song13,lu14,zhang14} Recently, Lu {\it et al.}\cite{lu14} succeeded in measuring the neutron scattering in the nematic phase of uniaxial strain-detwinned Ni-doped Ba-122 crystals, finding a transition of the low-energy spin excitations from four- to twofold symmetric at the same onset temperature as the in-plane resistivity anisotropy.\cite{tanatar10,chu10,ying11,chu12,blomberg13,ishida13,kuo14} 

In the low-temperature stripe ordered magnetic phase ($T < T_N$) of the 122 parent compounds, the spin response at low energies is naturally dominated by the steeply dispersing spin waves of the ordered state.\cite{lumsden10,harriger11,zhao08,ewings08,mcqueeney08,matan09,diallo09,zhao09,ewings11} Many studies have revealed the existence of three-dimensional strongly anisotropic spin waves emanating from $\mathbf{Q}_1$, and well-defined branches extending up to an energy scale of $\sim \!\!100$-$200$ meV. The fate of the spin waves at high energies is less settled and appears material dependent, some studies claim well-defined spin modes whereas others find highly damped excitations that eventually become ill-defined near the Brillouin zone boundary, presumably due to coupling with the particle-hole continuum.\cite{lumsden10,harriger11,zhao08,ewings08,mcqueeney08,matan09,diallo09,zhao09,ewings11}  Thus, the understanding of the spin dynamics in these systems remains a topic of significant interest, and whether a Heisenberg spin-only model with highly anisotropic exchange couplings or an itinerant Stoner-like spin density wave (SDW) scenario provides the most suitable description constitutes an important question that needs to be settled.\cite{lumsden10,mazin09,dai12}

The question of whether an itinerant or localized picture is more appropriate is intimately tied to the magnitude of the electronic correlations in these materials.\cite{si09,mannella14} At present this issue remains a research challenge, and simply highlights the importance of new detailed studies within each picture. It should be noted that recent dynamical mean field theory calculations (performed in the paramagnetic state) find that a description of the magnetic excitations requires both itinerant electrons and local moments with strong Hund's coupling.\cite{park11,liu12} Such studies also find a substantially larger mass enhancement of the iron selenides and iron chalcogenides compared to the iron pnictides, indicating that an itinerant approach may apply better to the latter systems.\cite{yin11} Even in the itinerant case, however, the theoretical description of the magnetic fluctuations is technically tedious due to five relevant Fe $3d$ orbitals and the existence of both intra- and inter-orbital Coulomb interactions.

Here, we present a general theoretical study of the magnetic excitations in the itinerant SDW phase within a five-orbital model relevant to the iron pnictides. The description is formulated in orbital space and includes therefore all orbitally dependent matrix element effects in the dynamical susceptibilities. The interaction consisting of the standard multi-orbital onsite Coulomb repulsion is included within all RPA bubble and ladder diagrams. Most earlier theoretical studies of spin excitations in the metallic SDW phase have been formulated within two- or three-band minimal models,\cite{brydon09,knolle10,knolle11} except from Ref.~\onlinecite{kaneshita10} where a five-band RPA calculation was used to obtain the dynamical susceptibilities with results in agreement with several experiments.\cite{harriger11,ewings11} Knolle {\it et al.} compared the exitonic and orbital scenarios in the simplified case of a two-orbital band.\cite{knolle11} Other theoretical studies have been performed within localized spin-only Heisenberg models,\cite{fang08,wang14,yao08,lv10,applegate10,conceicao11,wysocki11,goswami11} which capture e.g. the features of the spin-waves (Goldstone modes) but do not typically describe their damping due to electron-hole excitations which can be significant in metallic SDW systems. Our approach is similar to the approach of Ref.~\onlinecite{kaneshita10}, but we provide a comprehensive study of the crossover from broad diffuse spin excitations in the limit of weak interactions to sharp dispersive spin waves for larger interaction strengths. We find that the behavior of the high-energy spin modes depends qualitatively on the interaction parameters, which can be understood from an interaction-induced change of the directional-dependent damping. In addition, we resolve the orbital content of the spin waves, and discuss the evolution of spectral weight as a function of the parameters in our model. Finally, we also model the spin excitations in the paramagnetic nematic phase prior to entering the SDW phase. In this case, any finite orbital splitting of the $d_{xz}$ and $d_{yz}$ orbitals leads to, in principle, arbitrarily large spin anisotropy upon approaching the SDW instability (in temperature) from above, in qualitative agreement with recent neutron scattering measurements.\cite{lu14}

\section{model}

The starting point of the theoretical analysis is the following five-orbital Hamiltonian
\eq{
 	\mathcal{H}=\mathcal{H}_{0}+\mathcal{H}_{\text{int}} \ ,
}{eq:H}
where $\mathcal{H}_0$ constitutes the kinetic part obtained from a tight-binding fit to the DFT band-structure by Ikeda {\it et al.}\cite{ikeda10}
\eq{
	\mathcal{H}_{0}=\sum_{\mathbf{ij},\mu\nu,\sigma}t_{\mathbf{ij}}^{\mu\nu}c_{\mathbf{i}\mu\sigma}^{\dagger}c_{\mathbf{j}\nu\sigma}-\mu_0\sum_{\mathbf{i}\mu\sigma}n_{\mathbf{i}\mu\sigma}.
}{eq:H0}
Here, the operator $c_{\mathbf{i} \mu\sigma}^{\dagger}$ creates an electron at the $i$-th site in orbital $\mu$ with spin $\sigma$, and $\mu_0$ is the chemical potential which is fixed so that the doping $\delta=\langle n \rangle - 6.0 = 0.0$.
The indices $\mu$ and $\nu$ run through 1 to 5 corresponding to the Fe orbitals $d_{xz}$, $d_{yz}$, $d_{x^2-y^2}$, $d_{xy}$, and $d_{3z^2-r^2}$.

The second term in \eqref{eq:H} describes the Coulomb interaction restricted to intrasite processes
\eq{
 	\mathcal{H}_{\text{int}}=\ U\sum_{\mathbf{i},\mu}n_{\mathbf{i}\mu\uparrow}n_{\mathbf{i}\mu\downarrow}
			&+(U'-\frac{J}{2})\sum_{\mathbf{i},\mu<\nu,\sigma\sigma'}n_{\mathbf{i}\mu\sigma}n_{\mathbf{i}\nu\sigma'}
 \\
 	-2J\sum_{\mathbf{i},\mu<\nu}\mathbf{S}_{\mathbf{i}\mu}\cdot\mathbf{S}_{\mathbf{i}\nu}
	&+J'\sum_{\mathbf{i},\mu<\nu,\sigma}c_{\mathbf{i}\mu\sigma}^{\dagger}c_{\mathbf{i}\mu\bar{\sigma}}^{\dagger}c_{\mathbf{i}\nu\bar{\sigma}}c_{\mathbf{i}\nu\sigma}\ ,
}{eq:Hint}
which includes the intraorbital (interorbital) Hubbard interaction $U$ ($U'$), the Hund's rule coupling $J$, and the pair hopping energy $J'$.
We assume spin and orbitally rotation-invariant interactions $J'=J$ and $U'=U-2J$.

\subsection{The gap equations}

The magnetic order of the parent compounds of the iron pnictides is collinear with  $(\pi,0)$ ordering vector and an ordered moment aligned antiferromagnetically (ferromagnetically) along the $a$ axis ($b$ axis) of the orthorhombic lattice. The momentum-space version of the Hamiltonian \eqref{eq:Hint} when mean-field decoupled in terms of the fields 
\eq{
	n_{\mu\nu} &= \sum_{\mathbf{k} \sigma} \ev{\ocd{\mu\sigma}(\mathbf{k}) \oc{\nu\sigma}(\mathbf{k})} \ ,\\
	m_{\mu\nu} &= \sum_{\mathbf{k} \sigma} \ev{\ocd{\mu\sigma}(\mathbf{k}+\mathbf{Q}_1) \oc{\nu\sigma}(\mathbf{k})}\sgn\sigma \ ,
}{eq:}
is given by 
 \eq{
	\mathcal{H}_{\text{int}}^{\text{MF}} = \ \frac{1}{2N} \sum_{\mathbf{k}\sigma}  \sum_{\mu,\nu} &N_{\mu\nu}
	\ocd{ \mu \sigma}(\mathbf{k}) \oc{\nu \sigma}(\mathbf{k}) \ , \\
	+\frac{1}{2N} \sum_{\mathbf{k}\sigma}  \sum_{\mu,\nu} &M_{\mu\nu}
	\ocd{ \mu \sigma}(\mathbf{k}+\mathbf{Q}_1) \oc{\nu \sigma}(\mathbf{k}) \sgn\bar\sigma \ ,
}{eq:}
relevant for the SDW state with antiferromagnetic ordering vector $\mathbf{Q}_1=(\pi,0)$. Additionally, we assume $ \ev{\ocd{\mu\uparrow}(\mathbf{k}+\mathbf{Q}_1) \oc{\nu\uparrow}(\mathbf{k})} = -  \ev{\ocd{\mu\downarrow}(\mathbf{k}+\mathbf{Q}_1) \oc{\nu\downarrow}(\mathbf{k})}$.
The associated intra- and inter-orbital gap equations are given by the $5\times 5$ matrices $\hat{N}$ and $\hat{M}$ with elements
\eq{
	 N_{\mu\mu} &= U n_{\mu\mu} +(2U'-J) \sum_{\nu\neq\mu}  n_{\nu\nu} \ , \\
	N_{\mu\nu} &= J n_{\nu\mu} + J' n_{\mu\nu} -(U'-J) n_{\nu\mu}\ , 
}{eq:Nmv}
and 
\eq{
	M_{\mu\mu} =&U m_{\mu\mu} +  \sum_{\nu\neq\mu} Jm_{\nu\nu}\ , \\
	M_{\nu\mu}=&U'  m_{\mu\nu} + J' m_{\nu\mu} \ ,
}{eq:Mmv}
respectively, with $\mu \neq \nu$.
All the elements of $\hat{N}$ and $\hat{M}$, in addition to the chemical potential $\mu_0$, are determined self-consistently for a given set of interaction parameters $U$ and $J$ by diagonalization of the total SDW Hamiltonian written as a $10\times10$ matrix
\eq{
	\mathcal{H}^{\text{MF}}({\mathbf k},\sigma) = \left[\begin{array}{ccc} 
	\mathcal{H}_{0}({\mathbf k})+\hat{N} & \quad \hat{M}\sgn{\bar\sigma} \\ 
	& \\
	\hat{M}\sgn{\bar\sigma} &\quad \mathcal{H}_{0}({\mathbf k}+{\mathbf Q}_1)+ \hat{N}
	\end{array}\right] \ ,
}{h5SDWmatrix}
 in the basis $[\oc{1 \sigma}({\mathbf k}), \cdots, \oc{5 \sigma}({\mathbf k}), \oc{1 \sigma}({\mathbf k}+{\mathbf Q}_1), \cdots, \oc{5 \sigma}({\mathbf k}+{\mathbf Q}_1)]$  which couples $\mathbf{k}$ and $\mathbf{k}+\mathbf{Q}_1$ through $\hat{M}$ and renormalizes the chemical potential through $\hat{N}$.

\begin{figure}[t]
\includegraphics[width=0.99\columnwidth]{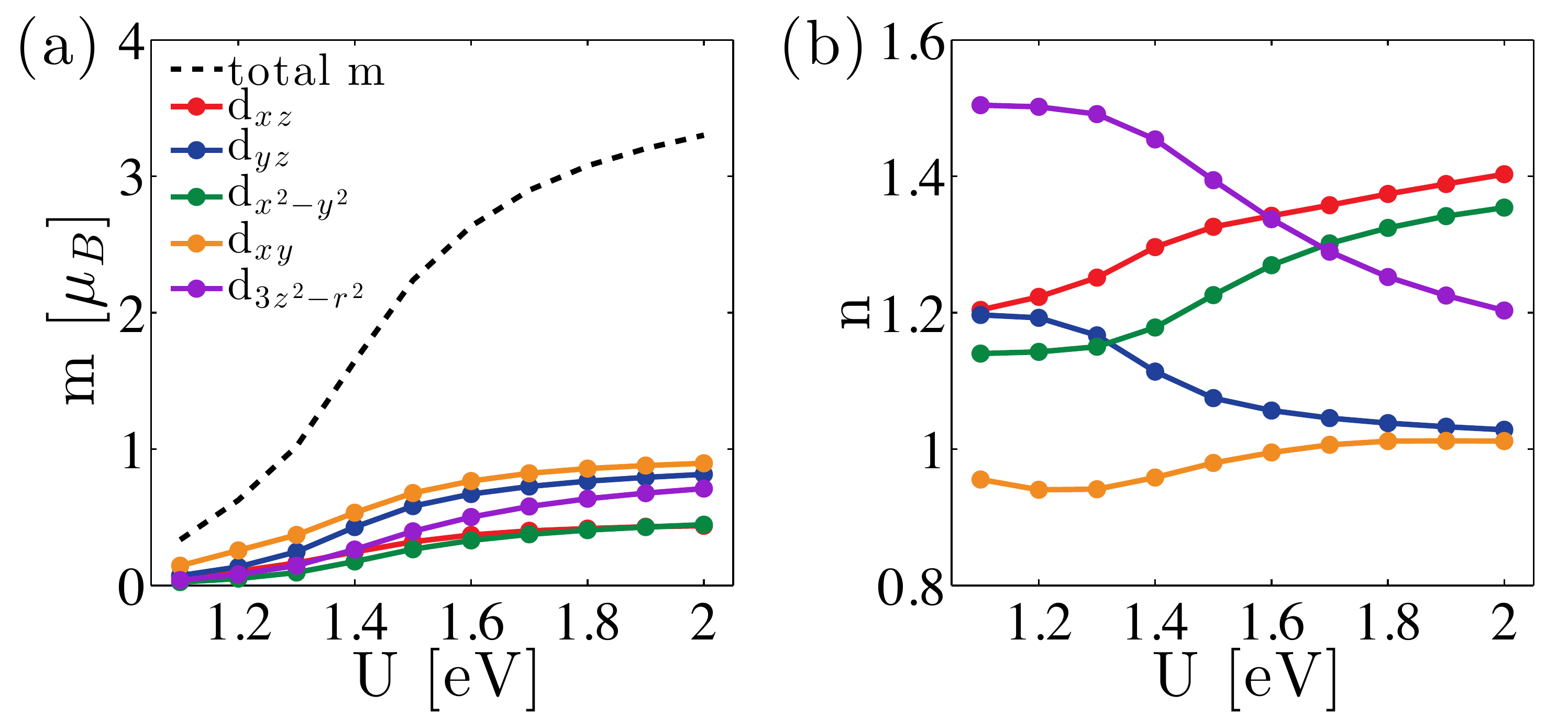}
\caption{(Color online) Self-consistent orbitally resolved magnetization (a) and electron density (b) versus interaction parameter $U$ for $J=U/6$.}
\label{fig:fields}
\end{figure}

In Fig.~\ref{fig:fields} we show the orbitally resolved magnetization and electron density as a function of Coulomb interaction $U$. The magnetic order induces an orbital splitting of the densities of the $d_{xz}$ and $d_{yz}$ orbitals. These results agree well with earlier studies of the static magnetic properties of the $(\pi,0)$ stripe phase.\cite{ccchen10,daghofer10,bascones12} We see that the $U$-range of Fig.~\ref{fig:fields} covers the interesting transition from very weak to almost saturated magnetization, hence in the following we focus on the evolution of the magnetic excitation spectrum in this same regime of $U$.

\subsection{The RPA spin susceptibility}
\begin{figure*}
\begin{tikzpicture}
\begin{scope}[scale=0.7] 
	\node at (-1,1.2) {(a)};
	\coordinate (L) at (3.5,0);
	\coordinate (w) at (1.5,0);
	\coordinate (A) at (0,0);
	\coordinate (A1) at (0,0);
	\coordinate (B1) at (1,1);
	\coordinate (B2) at (1,-1);
	\vertex{(A)};\draw (A) to[out=70,in=190] (B1);
	\vertex{(A)};\draw (A) to[out=-70,in=170] (B2);
	\coordinate (A) at ($(A) + (L)$);
	\coordinate (A2) at (A);
	\coordinate (C1) at ($(A) - (B2) - (w)$);
	\coordinate (C2) at  ($(A) - (B1) - (w)$);
	\coordinate (D1) at ($(A) - (B2)$);
	\coordinate (D2) at  ($(A) - (B1)$);
	\draw (B1) to node{$>$} (D1);\draw (B2) to node{$<$} (D2);
	\vertex{(A)};\draw (D1) to[out=-10,in=110] (A);
	\vertex{(A)};\draw (D2) to[out=10,in=-110] (A);
	\draw[thick] (C1) to (C2);\draw[thick] (C1) to (D1);
	\draw[thick] (D1) to (D2);\draw[thick] (C2) to (D2);
	\node at ($(A)- (w) - (0.2,0)$) {$\hat{\chi}^{+-}_{\text{RPA}}$};
	\node at ($(A1) + (0,1)$) {$\mu\uparrow$};\node at ($(A1)- (0,1)$) {$\nu\downarrow$};
	\node at ($(A2) + (0.1,1)$) {$\mu'\uparrow$};\node at ($(A2)- (0,1)$) {$\nu'\downarrow$};
	\node at ($(A2) + (1,0)$) {$=$};
	\coordinate (L) at (2.6,0);
	\coordinate (A) at ($(A) + (2,0)$);
	\coordinate (A1) at (A);
	\coordinate (B1) at ($ (1,1)$);
	\coordinate (B2) at ($ (1,-1)$);
	\vertex{(A)};\draw (A) to[out=70,in=185] ($(A1) +(B1)$);
	\vertex{(A)};\draw (A) to[out=-70,in=175] ($(A1) +(B2)$);
	\coordinate (A) at ($(A) + (L)$);
	\coordinate (A2) at (A);
	\coordinate (C1) at ($(A) - (B2) - (w)$);
	\coordinate (C2) at  ($(A) - (B1) - (w)$);
	\coordinate (D1) at ($(A) - (B2)$);
	\coordinate (D2) at  ($(A) - (B1)$);
	\draw ($(A1) +(B1)$) to node{$>$} (D1);\draw ($(A1) +(B2)$) to node{$<$} (D2);
	\vertex{(A)};\draw (D1) to[out=-5,in=110] (A);
	\vertex{(A)};\draw (D2) to[out=5,in=-110] (A);
	\node at ($(A1) + (0,1)$) {$\mu\uparrow$};\node at ($(A1)- (0,1)$) {$\nu\downarrow$};
	\node at ($(A2) + (0.1,1)$) {$\mu'\uparrow$};\node at ($(A2)- (0,1)$) {$\nu'\downarrow$};
	\node at ($(A2) + (1,0)$) {$+$};
	\coordinate (L) at (6.5,0);
	\coordinate (A) at ($(A) + (2,0)$);
	\coordinate (A1) at (A);
	\coordinate (B1) at ($ (1,1)$);
	\coordinate (B2) at ($ (1,-1)$);
	\vertex{(A)};\draw (A) to[out=70,in=190] ($(A1) +(B1)$);
	\vertex{(A)};\draw (A) to[out=-70,in=170] ($(A1) +(B2)$);
	\coordinate (B1b) at ($ (2,1)$);
	\coordinate (B2b) at ($ (2,-1)$);
	\draw[dashed] ($(A1) +(B1b)$) to node[right]{$\hat{g}^{\sigma\bar\sigma}$} ($(A1) +(B2b)$);
	\vertex{($(A1) +(B1b)$)};\vertex{($(A1) +(B2b)$)};
	\coordinate (A) at ($(A) + (L)$);
	\coordinate (A2) at (A);
	\coordinate (C1) at ($(A) - (B2) - (w)$);
	\coordinate (C2) at  ($(A) - (B1) - (w)$);
	\coordinate (D1) at ($(A) - (B2)$);
	\coordinate (D2) at  ($(A) - (B1)$);
	\draw ($(A1) +(B1)$) to ($(A1) +(B1b)$);\draw ($(A1) +(B2)$) to  ($(A1) +(B2b)$);
	\draw ($(A1) +(B1b)$) to node{$>$} (C1);\draw ($(A1) +(B2b)$) to node{$<$} (C2);
	\vertex{(A)};\draw (D1) to[out=-10,in=110] (A);
	\vertex{(A)};\draw (D2) to[out=10,in=-110] (A);
	\draw[thick] (C1) to (C2);\draw[thick] (C1) to (D1);
	\draw[thick] (D1) to (D2);\draw[thick] (C2) to (D2);
	\node at ($(A)- (w) - (0.2,0)$) {$\hat{\chi}^{+-}_{\text{RPA}}$};
	\node at ($(A1) + (0,1)$) {$\mu\uparrow$};\node at ($(A1)- (0,1)$) {$\nu\downarrow$};
	\node at ($(A2) + (0.1,1)$) {$\mu'\uparrow$};\node at ($(A2)- (0,1)$) {$\nu'\downarrow$};
	\node at ($(A1) + (B1b) + (-0.8,0.45)$) {$i \mathbf{\k}_1\uparrow$}  ; \node at ($(A1) +(B1b) + (0.8,0.45)$) {$j \mathbf{\k}_1' \uparrow $};
	\node at ($(A1) +(B2b) + (-0.8,-0.45)$) {$k  \mathbf{\k}_2 \downarrow $}; \node at ($(A1) +(B2b) + (0.8,-0.45)$) {$l\mathbf{\k}_2' \downarrow $};
\end{scope}
\begin{scope}[scale=0.7, yshift=-4cm] 
	\node at (-1,1.2) {(b)};
	\coordinate (L) at (3.5,0);
	\coordinate (w) at (1.5,0);
	\coordinate (A) at (0,0);
	\coordinate (A1) at (0,0);
	\coordinate (B1) at (1,1);
	\coordinate (B2) at (1,-1);
	\vertex{(A)};\draw (A) to[out=70,in=190] (B1);
	\vertex{(A)};\draw (A) to[out=-70,in=170] (B2);
	\coordinate (A) at ($(A) + (L)$);
	\coordinate (A2) at (A);
	\coordinate (C1) at ($(A) - (B2) - (w)$);
	\coordinate (C2) at  ($(A) - (B1) - (w)$);
	\coordinate (D1) at ($(A) - (B2)$);
	\coordinate (D2) at  ($(A) - (B1)$);
	\draw (B1) to node{$>$} (D1);\draw (B2) to node{$<$} (D2);
	\vertex{(A)};\draw (D1) to[out=-10,in=110] (A);
	\vertex{(A)};\draw (D2) to[out=10,in=-110] (A);
	\draw[thick] (C1) to (C2);\draw[thick] (C1) to (D1);
	\draw[thick] (D1) to (D2);\draw[thick] (C2) to (D2);
	\node at ($(A)- (w) - (0.2,0)$) {$\chi^{zz}_{\text{RPA}}$};
	\node at ($(A1) + (0,1)$) {$\mu\sigma$};\node at ($(A1)- (0,1)$) {$\nu\sigma$};
	\node at ($(A2) + (0.1,1)$) {$\mu'\sigma$};\node at ($(A2)- (0,1)$) {$\nu'\sigma$};
	\node at ($(A2) + (1,0)$) {$=$};
	\coordinate (L) at (2.6,0);
	\coordinate (A) at ($(A) + (2,0)$);
	\coordinate (A1) at (A);
	\coordinate (B1) at ($ (1,1)$);
	\coordinate (B2) at ($ (1,-1)$);
	\vertex{(A)};\draw (A) to[out=70,in=185] ($(A1) +(B1)$);
	\vertex{(A)};\draw (A) to[out=-70,in=175] ($(A1) +(B2)$);
	\coordinate (A) at ($(A) + (L)$);
	\coordinate (A2) at (A);
	\coordinate (C1) at ($(A) - (B2) - (w)$);
	\coordinate (C2) at  ($(A) - (B1) - (w)$);
	\coordinate (D1) at ($(A) - (B2)$);
	\coordinate (D2) at  ($(A) - (B1)$);
	\draw ($(A1) +(B1)$) to node{$>$} (D1);\draw ($(A1) +(B2)$) to node{$<$} (D2);
	\vertex{(A)};\draw (D1) to[out=-5,in=110] (A);
	\vertex{(A)};\draw (D2) to[out=5,in=-110] (A);
	\node at ($(A1) + (0,1)$) {$\mu\sigma$};\node at ($(A1)- (0,1)$) {$\nu\sigma$};
	\node at ($(A2) + (0.1,1)$) {$\mu'\sigma$};\node at ($(A2)- (0,1)$) {$\nu'\sigma$};
	\node at ($(A2) + (1,0)$) {$+$};
		\coordinate (L) at (2.6,0);
	\coordinate (A) at ($(A) + (2,0)$);
	\coordinate (A1) at (A);
	\coordinate (B1) at ($ (1,1)$);
	\coordinate (B2) at ($ (1,-1)$);
	\vertex{(A)};\draw (A) to[out=70,in=185] ($(A1) +(B1)$);
	\vertex{(A)};\draw (A) to[out=-70,in=175] ($(A1) +(B2)$);
	\coordinate (A) at ($(A) + (L)$);
	\coordinate (A2) at (A);
	\coordinate (C1) at ($(A) - (B2) - (w)$);
	\coordinate (C2) at  ($(A) - (B1) - (w)$);
	\coordinate (D1) at ($(A) - (B2)$);
	\coordinate (D2) at  ($(A) - (B1)$);
	\draw ($(A1) +(B1)$) to node{$>$} (D1);\draw ($(A1) +(B2)$) to node{$<$} (D2);
	\vertex{(A)};\draw (D1) to[out=-5,in=110] (A);
	\vertex{(A)};\draw (D2) to[out=5,in=-110] (A);
	\node at ($(A1) + (0,1)$) {$\mu\sigma$};\node at ($(A1)- (0,1)$) {$\nu \sigma$};
	\node at ($(A2) + (0.1,1)$) {$i \sigma'$};\node at ($(A2)- (0,1)$) {$j \sigma'$};
	\coordinate (L) at (2.6,0);
	\coordinate (A) at ($(A) + (2,0)$);
	\coordinate (A1) at (A);
	\coordinate (B1) at ($ (1,1)$);
	\coordinate (B2) at ($ (1,-1)$);
	\draw[dashed] (A1) -- node[below]{$g^{\sigma\bar\sigma}$}(A2);
	\vertex{(A)};\draw (A) to[out=70,in=185] ($(A1) +(B1)$);
	\vertex{(A)};\draw (A) to[out=-70,in=175] ($(A1) +(B2)$);
	\coordinate (L) at (3.5,0);
	\coordinate (A) at ($(A) + (L)$);
	\coordinate (A2) at (A);
	\coordinate (C1) at ($(A) - (B2) - (w)$);
	\coordinate (C2) at  ($(A) - (B1) - (w)$);
	\coordinate (D1) at ($(A) - (B2)$);
	\coordinate (D2) at  ($(A) - (B1)$);
	\draw ($(A1) +(B1)$) to node{$>$} (D1);\draw ($(A1) +(B2)$) to node{$<$} (D2);
	\vertex{(A)};\draw (D1) to[out=-5,in=110] (A);
	\vertex{(A)};\draw (D2) to[out=5,in=-110] (A);
	\node at ($(A1) + (0,1)$) {$k \bar\sigma'$};\node at ($(A1)- (0,1)$) {$l \bar\sigma'$};
	\node at ($(A2) + (0.1,1)$) {$\mu'\sigma$};\node at ($(A2)- (0,1)$) {$\nu'\sigma$};
	\draw[thick] (C1) to (C2);\draw[thick] (C1) to (D1);
	\draw[thick] (D1) to (D2);\draw[thick] (C2) to (D2);
	\node at ($(A)- (w) - (0.2,0)$) {$\chi^{zz}_{\text{RPA}}$};
	\coordinate (L) at (5,0);
	\coordinate (A) at ($ (6.5,-3.5)$);
	\coordinate (A1) at (A);
	\coordinate (B1) at ($ (1,1)$);
	\coordinate (B2) at ($ (1,-1)$);
		\node at ($(A) - (1,0)$) {$+$};
	\vertex{(A)};\draw (A) to[out=70,in=190] ($(A1) +(B1)$);
	\vertex{(A)};\draw (A) to[out=-70,in=170] ($(A1) +(B2)$);
	\coordinate (B1b) at ($ (1.2,1)$);
	\coordinate (B2b) at ($ (1.2,-1)$);
	\draw[snake] ($(A1) +(B1b)$) to node[right]{$g^{\sigma\sigma}$} ($(A1) +(B2b)$);
	\vertex{($(A1) +(B1b)$)};\vertex{($(A1) +(B2b)$)};
	\coordinate (A) at ($(A) + (L)$);
	\coordinate (A2) at (A);
	\coordinate (C1) at ($(A) - (B2) - (w)$);
	\coordinate (C2) at  ($(A) - (B1) - (w)$);
	\coordinate (D1) at ($(A) - (B2)$);
	\coordinate (D2) at  ($(A) - (B1)$);
	\draw ($(A1) +(B1)$) to ($(A1) +(B1b)$);\draw ($(A1) +(B2)$) to  ($(A1) +(B2b)$);
	\draw ($(A1) +(B1b)$) to node{$>$} (C1);\draw ($(A1) +(B2b)$) to node{$<$} (C2);
	\vertex{(A)};\draw (D1) to[out=-10,in=110] (A);
	\vertex{(A)};\draw (D2) to[out=10,in=-110] (A);
	\draw[thick] (C1) to (C2);\draw[thick] (C1) to (D1);
	\draw[thick] (D1) to (D2);\draw[thick] (C2) to (D2);
	\node at ($(A)- (w) - (0.2,0)$) {$\chi^{zz}_{\text{RPA}}$};
	\node at ($(A1) + (0,1)$) {$\mu\sigma$};\node at ($(A1)- (0,1)$) {$\nu\sigma$};
	\node at ($(A2) + (0.1,1)$) {$\mu'\sigma$};\node at ($(A2)- (0,1)$) {$\nu'\sigma$};
	\node at ($(A1) + (B1b) + (-0.3,0.45)$) {$i$}; \node at ($(A1) +(B1b) + (0.3,0.45)$) {$j$};
	\node at ($(A1) +(B2b) + (-0.3,-0.45)$) {$k$}; \node at ($(A1) +(B2b) + (0.3,-0.45)$) {$l$};
	\node at ($(A2) + (1,0)$) {$+$};
		\coordinate (L) at (2.6,0);
	\coordinate (A) at ($(A) + (2,0)$);
	\coordinate (A1) at (A);
	\coordinate (B1) at ($ (1,1)$);
	\coordinate (B2) at ($ (1,-1)$);
	\vertex{(A)};\draw (A) to[out=70,in=185] ($(A1) +(B1)$);
	\vertex{(A)};\draw (A) to[out=-70,in=175] ($(A1) +(B2)$);
	\coordinate (A) at ($(A) + (L)$);
	\coordinate (A2) at (A);
	\coordinate (C1) at ($(A) - (B2) - (w)$);
	\coordinate (C2) at  ($(A) - (B1) - (w)$);
	\coordinate (D1) at ($(A) - (B2)$);
	\coordinate (D2) at  ($(A) - (B1)$);
	\draw ($(A1) +(B1)$) to node{$>$} (D1);\draw ($(A1) +(B2)$) to node{$<$} (D2);
	\vertex{(A)};\draw (D1) to[out=-5,in=110] (A);
	\vertex{(A)};\draw (D2) to[out=5,in=-110] (A);
	\node at ($(A1) + (0,1)$) {$\mu\sigma$};\node at ($(A1)- (0,1)$) {$\nu \sigma$};
	\node at ($(A2) + (0.1,1)$) {$i \sigma'$};\node at ($(A2)- (0,1)$) {$j \sigma'$};
	\coordinate (L) at (2.6,0);
	\coordinate (A) at ($(A) + (2,0)$);
	\coordinate (A1) at (A);
	\coordinate (B1) at ($ (1,1)$);
	\coordinate (B2) at ($ (1,-1)$);
	\draw[snake] (A1) -- node[below]{$g^{\sigma\sigma}$}(A2);
	\vertex{(A)};\draw (A) to[out=70,in=185] ($(A1) +(B1)$);
	\vertex{(A)};\draw (A) to[out=-70,in=175] ($(A1) +(B2)$);
	\coordinate (L) at (3.5,0);
	\coordinate (A) at ($(A) + (L)$);
	\coordinate (A2) at (A);
	\coordinate (C1) at ($(A) - (B2) - (w)$);
	\coordinate (C2) at  ($(A) - (B1) - (w)$);
	\coordinate (D1) at ($(A) - (B2)$);
	\coordinate (D2) at  ($(A) - (B1)$);
	\draw ($(A1) +(B1)$) to node{$>$} (D1);\draw ($(A1) +(B2)$) to node{$<$} (D2);
	\vertex{(A)};\draw (D1) to[out=-5,in=110] (A);
	\vertex{(A)};\draw (D2) to[out=5,in=-110] (A);
	\node at ($(A1) + (0,1)$) {$k \sigma'$};\node at ($(A1)- (0,1)$) {$l \sigma'$};
	\node at ($(A2) + (0.1,1)$) {$\mu'\sigma$};\node at ($(A2)- (0,1)$) {$\nu'\sigma$};
	\draw[thick] (C1) to (C2);\draw[thick] (C1) to (D1);
	\draw[thick] (D1) to (D2);\draw[thick] (C2) to (D2);
	\node at ($(A)- (w) - (0.2,0)$) {$\chi^{zz}_{\text{RPA}}$};
\end{scope}
\end{tikzpicture}
\caption{(a) The diagrammatic RPA series of the transverse spin susceptibility in the SDW state where the propagators correspond to the full Greens functions in orbital and  $\mathbf{q}$,$\mathbf{q'}$ space and $\hat{g}^{\sigma\bar\sigma} =g^{\sigma\bar\sigma}  \delta_{\mathbf{\k}_1 + \mathbf{\k}_2, \mathbf{\k}_1' + \mathbf{\k}_2' }$ is a diagonal matrix in $\mathbf{q}$,$\mathbf{q'}$ space. The paramagnetic normal state expression is obtained by considering normal propagators only.  (b) The diagrammatic RPA series of the longitudinal spin susceptibility with both bubble and ladder type diagrams. The same expression applies to both the normal and the SDW phase in the longitudinal channel because the Umklapp processes do not contribute in this case.}
\label{fig:diagrams}
\end{figure*}
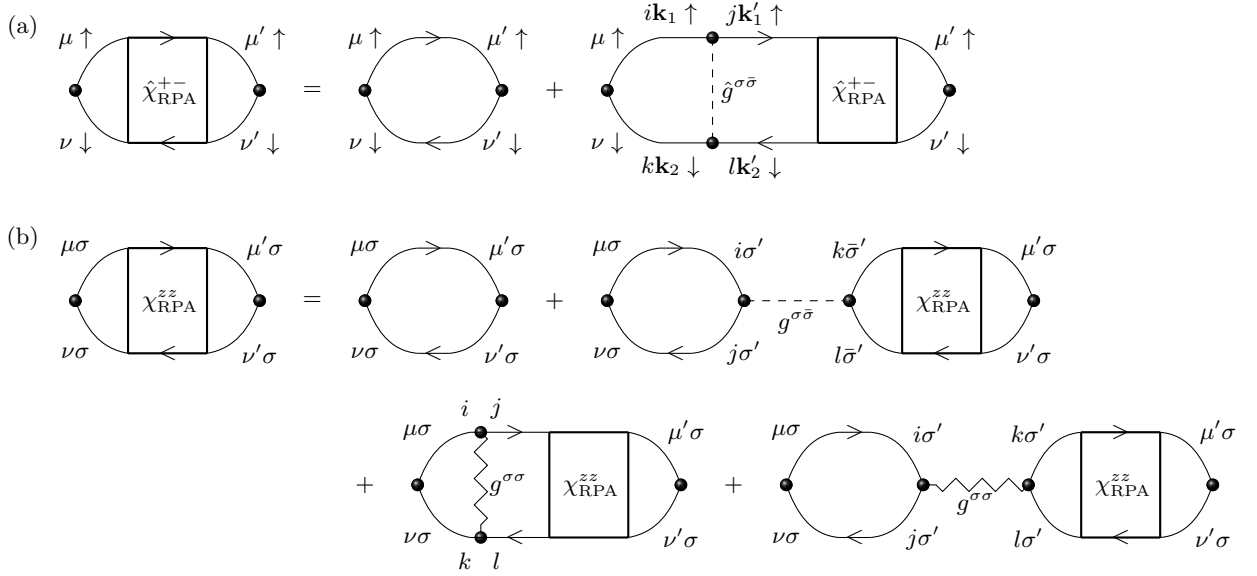

To investigate the collective spin excitations we compute the imaginary part of the physical spin susceptibility given by 
\begin{equation}\label{chi}
	\chi^{lm} (\mathbf{q}, i \omega_n) \equiv \frac{1}{2}\sum_{sp}[\chi^{lm}]^{sp}_{sp}(\mathbf{q}, \mathbf{q}, i \omega_n),
\end{equation}
with
\eq{
[\chi^{lm}]^{\mu\mu'}_{\nu\nu'}(\mathbf{q}, \mathbf{q'}, i\omega_n) = & -\frac{1}{\mathcal{N}} \int_0^\beta d\tau e^{i\omega_n \tau} \\
&\quad \times \evt{S_{\mu'\nu'}^{l} (\mathbf{q'};\tau) S_{\nu\mu}^{m} (-\mathbf{q})} \ ,
}{chiqqprime}
where $S^{i}_{\alpha \beta}$ denotes the (generalized) spin operator $i$ with an electron (hole) in orbital $\alpha$ ($\beta$). 

In the current case of a stripe-order SDW, the spin susceptibilities of \eqref{chiqqprime} become $2 \times 2$ matrices $\hat{\chi}^{lm}(\mathbf{q}, \omega)$ in $\mathbf{q}, \mathbf{q}'$ space where the 11-component  ($\mathbf{q} = \mathbf{q'}$) corresponds to the physical contribution, and the off-diagonal elements ($\mathbf{q'}=\mathbf{q} \pm \mathbf{Q}_1$) account for Umklapp scattering between neighboring reduced Brillouin zones (RBZ).\cite{brydon09,knolle11} The transverse non-interacting spin susceptibility matrix is given by 
\eq{
	&[\hat{\chi}^{+-}_0]^{\mu\mu'}_{\nu\nu'}(\mathbf{q}, i\omega_n) 
	=\left[\begin{array}{cc} 
	\chi^{+-,0}_{00} & \chi^{+-,0}_{0Q_1}  \\
	\chi^{+-,0}_{Q_10}  & \chi^{+-,0}_{Q_1Q_1}
	\end{array}\right]^{\mu\mu'}_{\nu\nu'} \  \\
	\!&=\!
	\sum_{\mathbf{k} \in \text{RBZ}} \sum_{n,m =1}^{10}  
	 \left[\begin{array}{cc} 
	\mathcal{M}^{00}_{nm}({\mathbf k},{\mathbf q}) &
	 \mathcal{M}^{0Q_1}_{nm}({\mathbf k},{\mathbf q})
	\\
	\mathcal{M}^{Q_10}_{nm}({\mathbf k},{\mathbf q}) & \mathcal{M}^{Q_1Q_1}_{nm}({\mathbf k},{\mathbf q}) \end{array}\right]^{\mu\mu'}_{\nu\nu'} 
	\\
	&\quad \quad \quad\quad \quad\quad  \times\ 
	\frac{f(E_{m\mathbf{k}+\mathbf{q}}) - f(E_{n\mathbf{k}})}{i\omega_n - E_{m\mathbf{k}+\mathbf{q}} + E_{n\mathbf{k}}}\ ,
}{eq:chi00}
where we use the shorthand notation $\chi^{lm,0}_{ab} \equiv \chi^{lm}_{0}(\mathbf{q}+\mathbf{a}, \mathbf{q}+\mathbf{b}; i\omega_n)$ and specify the matrix elements $[\mathcal{M}^{qq'}_{nm}({\mathbf k},{\mathbf q})] ^{\mu\mu'}_{\nu\nu'}$ in Table \ref{coferenceFactors}. A diagrammatic inspection of Table~\ref{coferenceFactors} additionally shows that each of the bare matrix elements gives rise to four pair bubble diagrams in $\mathbf{q, q'}$ space composed of normal and Umklapp propagators with the diagonal (off-diagonal) pair bubbles consisting of an even (odd) number of Umklapp propagators. 

\begin{table}[b]
\caption{The matrix elements of the SDW state presented in terms of $\mathcal{A}^{\mu\mu'}_{\nu\nu'} = (a_{n\mathbf{k}\downarrow}^{\nu'})^* a_{n\mathbf{k}\downarrow}^{\nu} (a_{m\mathbf{k}+\mathbf{q}\uparrow}^{\mu})^* a_{m\mathbf{k}+\mathbf{q}\uparrow}^{\mu'}$, where $a_{n\mathbf{k}\sigma}^{\alpha}=a_{n\mathbf{k}\sigma}^{\alpha + 10} $ is the element $\alpha$ of the band eigenvector $n$ of $\mathcal{H}^{\text{MF}}({\mathbf k},\sigma)$.}
\begin{center}
\begin{tabular}{|l|c|}
\hline
\hline
 $[\mathcal{M}^{00}_{nm}({\mathbf k},{\mathbf q})]^{\mu\mu'}_{\nu\nu'}  $ & $\mathcal{A}^{\mu\mu'}_{\nu\nu'} + \mathcal{A}^{\mu , \mu'+5}_{\nu , \nu'+5} + \mathcal{A}^{\mu+5, \mu'}_{\nu+5, \nu'} + \mathcal{A}^{\mu+5, \mu'+ 5}_{\nu+5, \nu'+ 5}$ \\
 $[\mathcal{M}^{Q_1Q_1}_{nm}({\mathbf k},{\mathbf q})]^{\mu\mu'}_{\nu\nu'}$ &	$[\mathcal{M}^{00}_{nm}({\mathbf k},{\mathbf q})]^{\mu, \mu'+5}_{\nu+5, \nu'}$ \\
 $[\mathcal{M}^{0Q_1}_{nm}({\mathbf k},{\mathbf q})]^{\mu\mu'}_{\nu\nu'}$ &		$[\mathcal{M}^{00}_{nm}({\mathbf k},{\mathbf q})]^{\mu, \mu'+5}_{\nu, \nu'}$ \\
  $[\mathcal{M}^{Q_1 0}_{nm}({\mathbf k},{\mathbf q})]^{\mu\mu'}_{\nu\nu'}$ & 	$[\mathcal{M}^{00}_{nm}({\mathbf k},{\mathbf q})]^{\mu, \mu'}_{\nu+5, \nu'}$\\ 
  \hline
\hline
\end{tabular}
\end{center}
\label{coferenceFactors}
\end{table}%

To obtain the desired RPA expression for the physical 11-component $\chi^{+-}_{\text{RPA}}(\mathbf{q}, i\omega_n) \equiv \chi^{+-, \text{RPA}}_{00}$, we sum the standard RPA ladder and bubble diagrams by introducing the Coulomb interactions through the interaction matrix $g^{\sigma\sigma'}$ with the following non-zero elements
\eq{
	[g^{\sigma\bar\sigma} ]^{\mu\mu}_{\mu\mu} &= U\ , \quad \quad [g^{\sigma\bar\sigma} ]^{\mu \mu}_{\nu \nu} = U' = U - 2J\ , \\
	[g^{\sigma\bar\sigma} ]^{\mu \nu}_{\nu \mu} &= J\ , \quad \quad [g^{\sigma\bar\sigma} ]^{\mu \nu}_{\mu \nu} = J' = J \ , 
}{gopposite}
and 
\eq{
	[g^{\sigma\sigma} ]^{\mu \nu}_{\nu \mu} &= U' - J\ .
}{gsame}

In Fig.~\ref{fig:diagrams}(a) we depict the RPA series of the transverse channel in terms of full Greens functions with a matrix structure in both orbital and $\mathbf{q, q'}$ space. The latter is important because we have non-zero Umklapp components in the non-interacting transverse spin susceptibility of \eqref{eq:chi00}. Thus, the obtained diagrammatic series consists solely of ladder type diagrams connected by opposite-spin interaction processes, \eqref{gopposite}, occurring locally within one RBZ. 

With this we arrive at the following Dyson-like equation for the total transverse RPA spin susceptibility in the symmetry-broken SDW state 
\begin{widetext} 
\eq{
	[\hat{\chi}^{+-}_{\text{RPA}}]^{\mu\mu'}_{\nu\nu'}=\left[\begin{array}{cc}
	\chi^{+-,0}_{00}  & \chi^{+-,0}_{0Q_1} 
	\\
	 \chi^{+-,0}_{Q_10}  & \chi^{+-,0}_{Q_1Q_1} \end{array}\right]^{\mu\mu'}_{\nu\nu'}
	 \!\!+ \sum_{ijkl}
	 \left[\begin{array}{cc}
	\chi^{+-,0}_{00}  & \chi^{+-,0}_{0Q_1} 
	\\
	 \chi^{+-,0}_{Q_10}  & \chi^{+-,0}_{Q_1Q_1} \end{array}\right]^{\mu i}_{\nu k} 
	  \left[\begin{array}{cc}
	 [g^{\sigma\bar\sigma}]^{ij}_{kl}   & 0
	\\
	0  & [g^{\sigma\bar\sigma}]^{ij}_{kl}  \end{array}\right]
	 \left[\begin{array}{cc}
	\chi^{+-,\text{RPA}}_{00}  & \chi^{+-,\text{RPA}}_{0Q_1} 
	\\
	 \chi^{+-,\text{RPA}}_{Q_10}  & \chi^{+-,\text{RPA}}_{Q_1Q_1} \end{array}\right]^{j \mu'}_{l \nu'} \ ,
	}{eq:trans_RPA}
	\end{widetext}
showing explicitly that the Umklapp components contribute to the physical $\chi^{+-,\text{RPA}}_{00}$. 

By contrast, the non-interacting Umklapp components vanish in the longitudinal channel but the longitudinal RPA expression is further complicated by the mixing of bubble and ladder diagrams since it is possible to generate diagrams using all the interactions processes of \eqref{gopposite} and \eqref{gsame}. The diagrammatic RPA series of the longitudinal channel is shown in Fig.~\ref{fig:diagrams}(b). The longitudinal RPA spin susceptibility is not affected by the symmetry-breaking of the SDW state and becomes simply
\eq{
	[\chi_{\text{RPA}}^{zz}]^{\mu\mu'}_{\nu\nu'} &= [\chi_{0}^{zz}]^{\mu\mu'}_{\nu\nu'} + \sum_{ijkl}  [\chi_{0}^{zz}]^{\mu i}_{\nu k}
	 [g^{zz}]^{i j}_{k l}  [\chi_{\text{RPA}}^{zz}]^{j \mu'}_{l \nu'}, \\
	  [g^{zz}]^{i j}_{k l} &=  [g^{\sigma\sigma}]^{i j}_{k l} - [g^{\sigma\sigma}]^{i k}_{j l} + [g^{\sigma\bar\sigma}]^{i k}_{j l} ,
}{}
which is obtained from combining the geometrical series given by
\begin{widetext}
\eq{
	\left[\begin{array}{c}\chi_{\text{RPA}}^{\uparrow\uparrow} \\ \chi_{\text{RPA}}^{\uparrow\downarrow}\end{array}\right]^{\mu\mu'}_{\nu\nu'}
	=
	\left[\begin{array}{c}\chi_{0}^{\uparrow\uparrow} \\ 0\end{array}\right]^{\mu\mu'}_{\nu\nu'}
	+ \sum_{jl}
	 \left[\begin{array}{cc}
	 [\chi_{0}^{\uparrow\uparrow}]^{\mu i}_{\nu k} \{ [g^{\sigma\sigma}]^{ij}_{kl} - [g^{\sigma\sigma}]^{ik}_{jl} \}			
	 &  -  [\chi_{0}^{\uparrow\uparrow}]^{\mu i}_{\nu k}  [g^{\sigma\bar\sigma}]^{ik}_{jl} 
	 \\
	-  [\chi_{0}^{\downarrow\downarrow}]^{\mu i}_{\nu k}  [g^{\sigma\bar\sigma}]^{ik}_{jl} 
	& 
	 [\chi_{0}^{\downarrow\downarrow}]^{\mu i}_{\nu k} \{ [g^{\sigma\sigma}]^{ij}_{kl} -  [g^{\sigma\sigma}]^{ik}_{jl} \}
	\end{array}\right] 
	\left[\begin{array}{c}\chi_{\text{RPA}}^{\uparrow\uparrow} \\ \chi_{\text{RPA}}^{\uparrow\downarrow}\end{array}\right]^{\mu\mu'}_{\nu\nu'} \ .
}{}
\end{widetext}

As a consistency check, we have verified the expected spin-rotational invariance of the paramagnetic state when setting the Umklapp elements of the transverse spin susceptibility to zero, and assuming $J=J'$ since then
\eq{
	  [g^{zz}]^{\mu \mu }_{\mu \mu} &= U = [g^{\sigma\bar\sigma}]^{\mu \mu }_{\mu \mu}, \\
	   [g^{zz}]^{\mu \mu }_{\nu \nu} &= U' - J + J' =[g^{\sigma\bar\sigma}]^{\mu \mu }_{\nu \nu},\\
	   [g^{zz}]^{\mu \nu }_{\nu \mu} &= J =  [g^{\sigma\bar\sigma}]^{\mu \nu }_{\nu \mu}, \\
	    [g^{zz}]^{\mu \nu }_{\mu \nu} &=  -U' + J + U' = J =  [g^{\sigma\bar\sigma}]^{\mu \nu }_{\nu \mu}  \ .
}{} 

In the following we focus the discussion of the results on the transverse dynamical spin susceptibility obtained from solving \eqref{eq:trans_RPA} and subsequently computing the full susceptibility given by \eqref{chi} since this is the main observable relevant to neutron scattering experiments.


\section{results}

In the following we show the results from evaluating the RPA dynamical susceptibility numerically in the SDW and paramagnetic nematic phases, respectively. For the results relevant to the SDW (nematic) phase, we have summed over a $100\times 100$ ($200\times 200$) $k$-mesh in \eqref{eq:chi00} and used a small artificial smearing $\eta = 1$ ($\eta = 3$) meV in $i\omega_n \rightarrow \omega + i\eta$. For clarity we display all results for a fixed antiferromagnetic (ferromagnetic) axis corresponding to a single orthorhombic domain or a fully detwinned sample.

\subsection{Spin excitations in the SDW phase}

\begin{figure}[t]
\includegraphics[width=0.99\columnwidth]{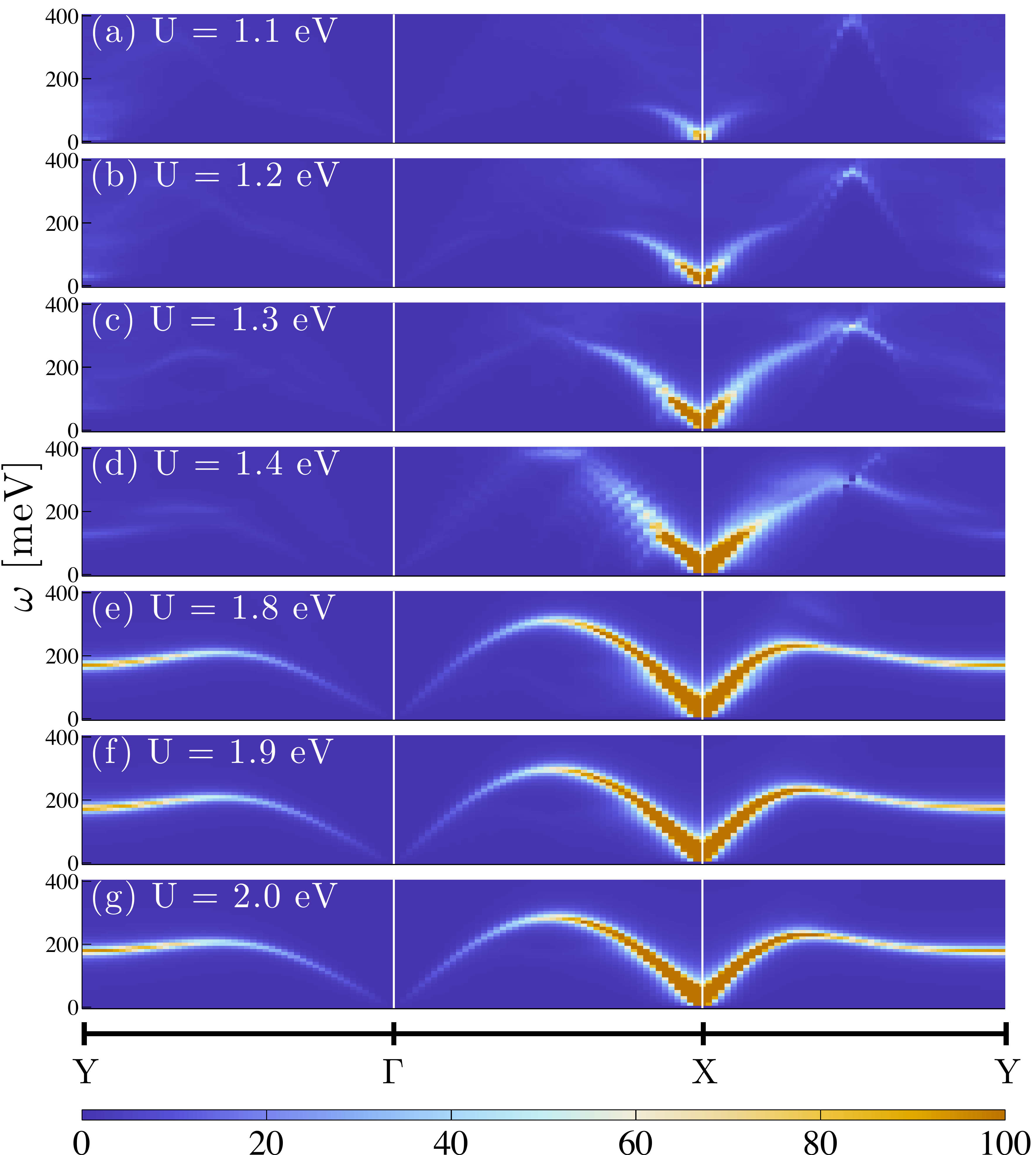}
\caption{(Color online) Imaginary part of the physical transverse dynamical RPA susceptibility, $\mbox{Im} \chi^{+-}_{\text{RPA}}({\mathbf q},\omega)$, in the SDW phase along the momentum space path $Y \rightarrow \Gamma \rightarrow X \rightarrow Y$ for different values of $U=1.1, 1.2, 1.3, 1.4, 1.8, 1.9, 2.0$ eV (top to bottom) shown in each panel.}
\label{fig:dispvsU}
\end{figure}

We begin the discussion of the results by presenting the physical dynamical transverse spin susceptibility in the ordered SDW phase. We note that within the formalism of the present paper, a high degree of convergence and the inclusion of the inter-orbital density terms $n_{\nu\mu}$ are crucial for fulfilling Goldstone's theorem and obtaining reliable results. In Fig.~\ref{fig:dispvsU} we show the momentum and energy dependence of the imaginary part of $\chi^{+-}_{\text{RPA}}({\mathbf q},\omega)$ at $T \ll T_N$ for increasing interaction $U$ with a fixed ratio $J=U/6$. The momentum axis follows the closed triangular path $Y \rightarrow \Gamma \rightarrow X \rightarrow Y$ with $Y=(0,\pi)$, $\Gamma=(0,0)$, and $X=(\pi,0)$. From Fig.~\ref{fig:dispvsU} we clearly see how the broad diffusive spin waves pivoted at $X$ in the low-$U$ limit sharpen up and get protected by the growing local SDW gap to the particle-hole continuum as $U$ is increased. By contrast, at low $U$ the spin modes are generally overdamped and become dissolved by the continuum already at low energies. At $Y$ there are no low-energy collective modes, as opposed to the result within the excitonic scenario,\cite{knolle11} but rather a high-energy branch in the large $U$ limit. By comparison to Fig.~\ref{fig:fields}(a), it is evident that the evolution of the spin dynamics in Fig.~\ref{fig:dispvsU} traces directly the enhanced magnetization as expected from a Stoner scenario. Finally, note from Fig.~\ref{fig:dispvsU} that the imaginary part of $\chi^{+-}_{\text{RPA}}({\mathbf q},\omega)$ always exhibits zero-energy modes simply because we do not include magnetic anisotropy in the model.

\begin{figure}[t]
\includegraphics[width=0.99\columnwidth]{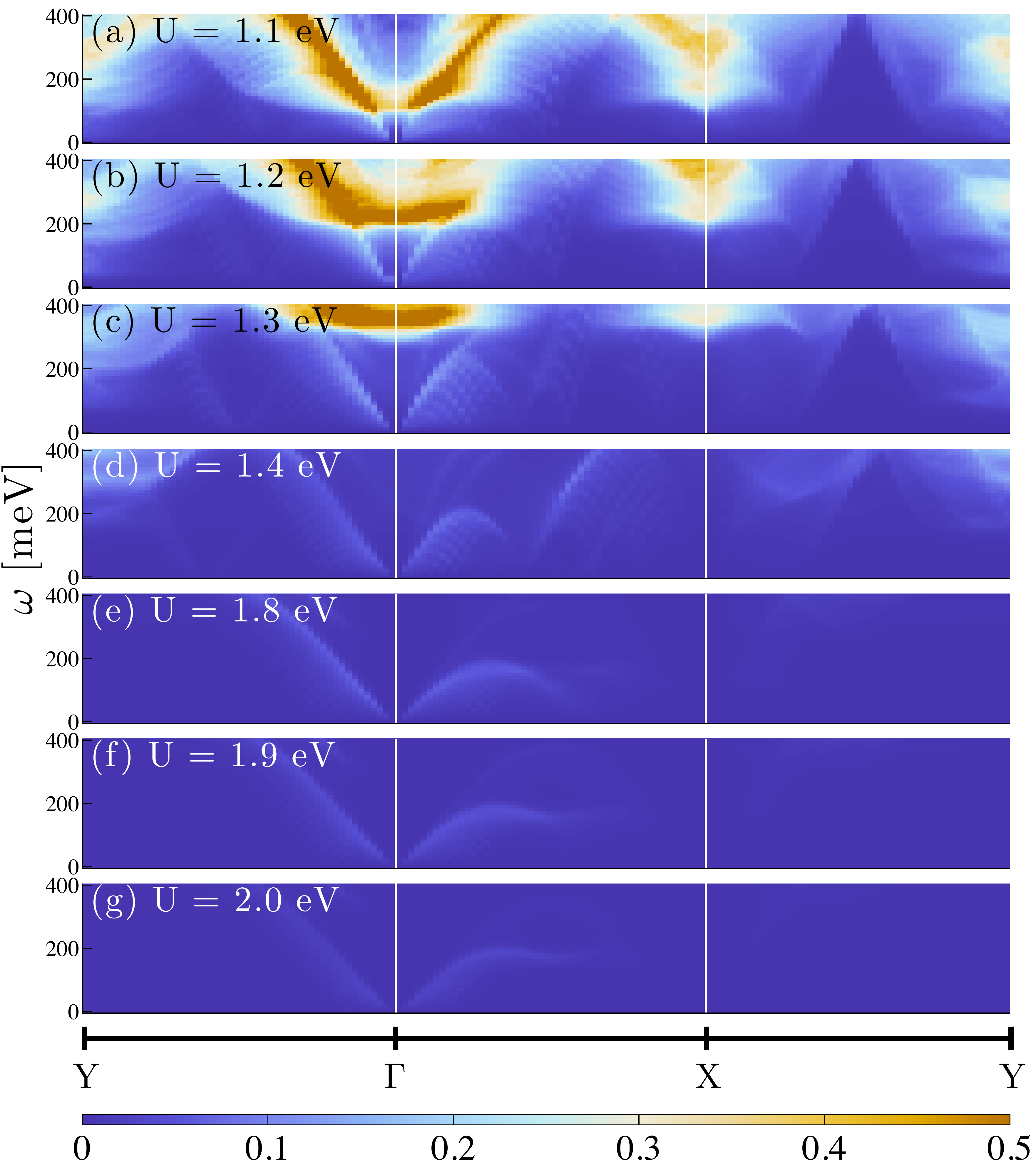}
\caption{(Color online) Imaginary part of the bare physical transverse spin susceptibility, $\mbox{Im}\chi^{+-}_{0}({\mathbf q},\omega)$, for the same values of $U$ and momenta as in Fig.~\ref{fig:dispvsU}.}
\label{fig:dispvsU0}
\end{figure}

 \begin{figure}[hb!]
\includegraphics[width=0.99\columnwidth]{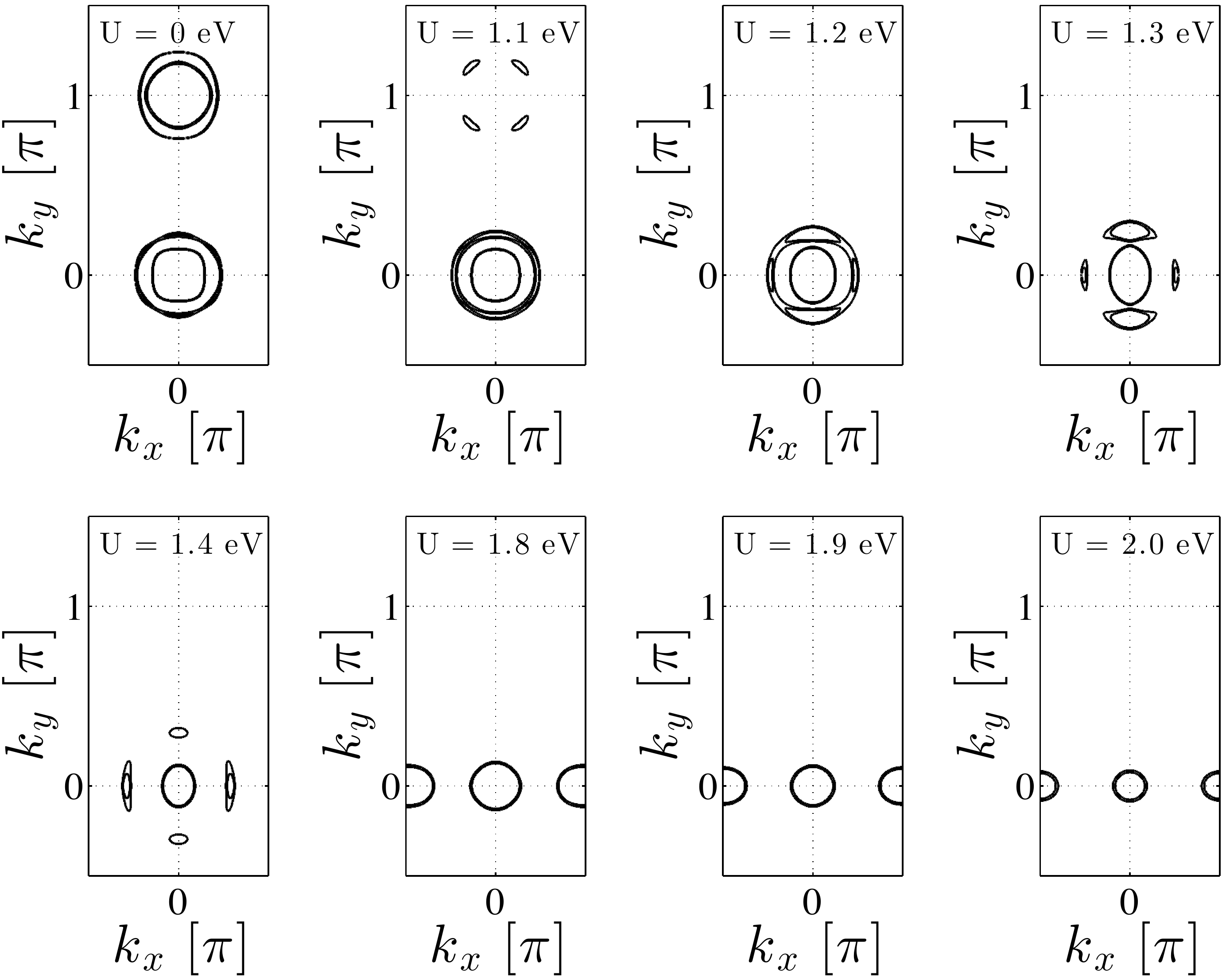}
\caption{Evolution of the Fermi surface as a function of the SDW order parameter for the values of Coulomb $U$ stated in the insets. Each panel shows the spectral weight in a window of $\pm 5$ meV around the Fermi level.}
 \label{fig:FS}
\end{figure}
 
\begin{figure}[b]
\includegraphics[height=0.43\columnwidth]{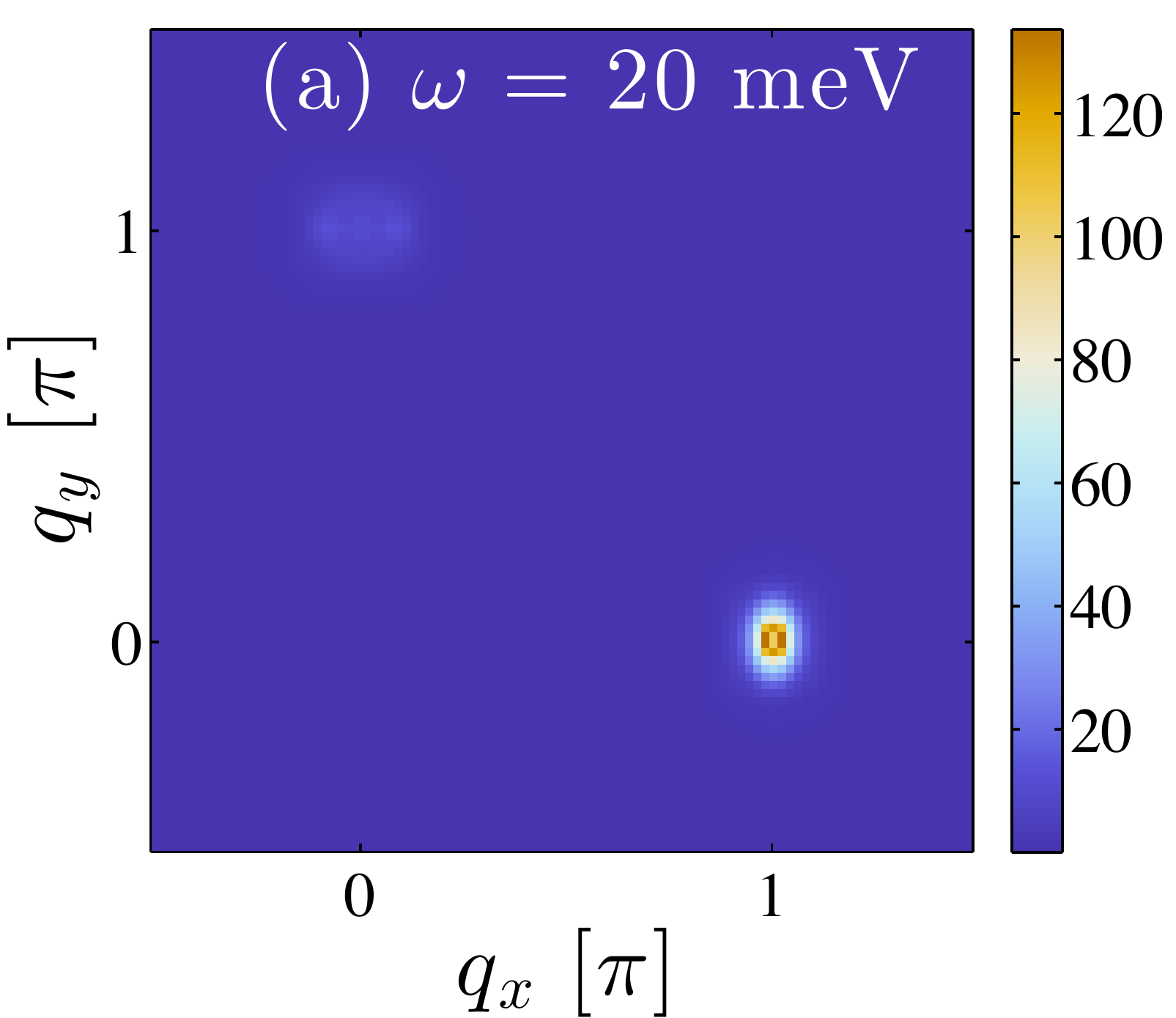}
\includegraphics[height=0.43\columnwidth]{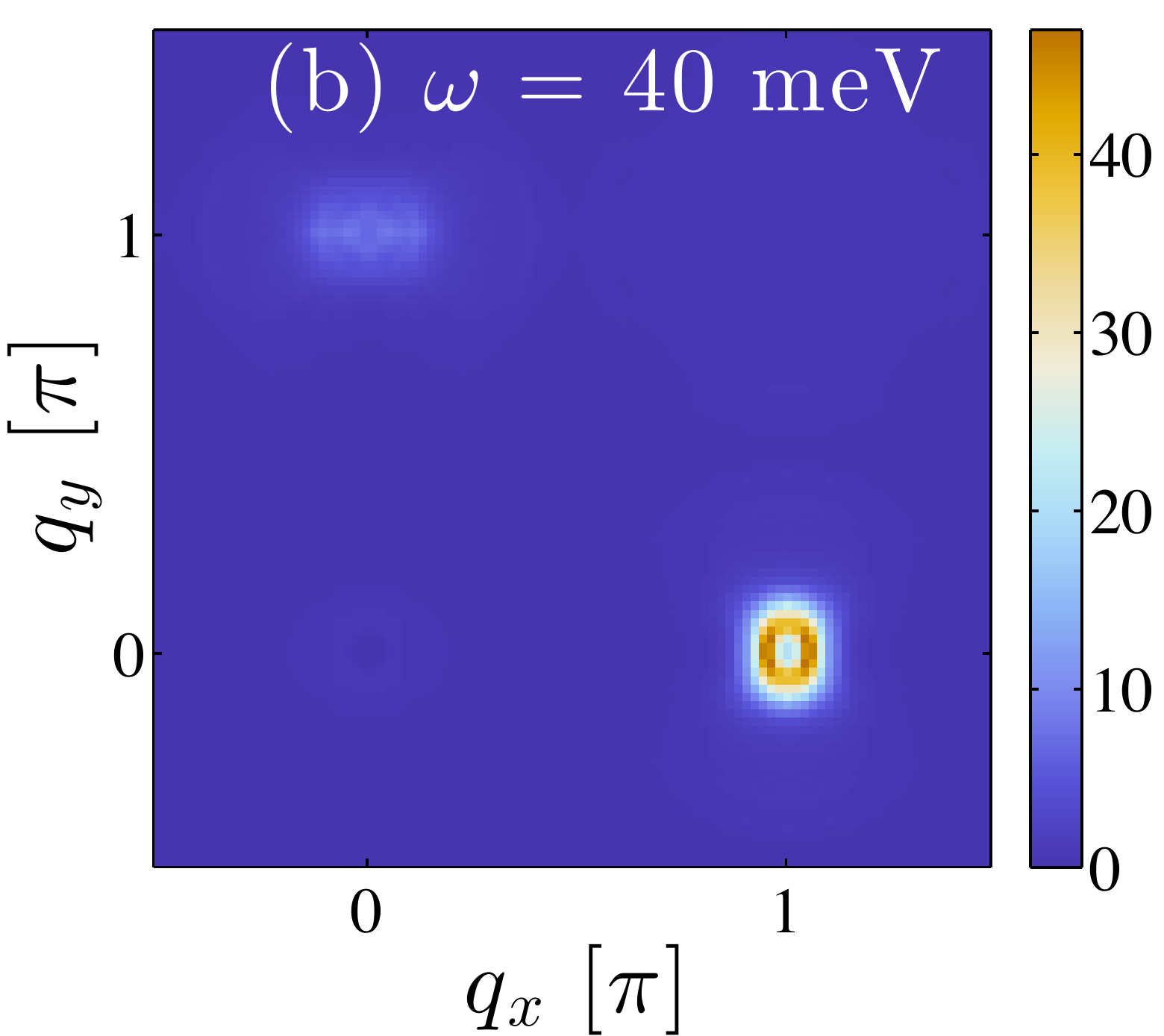}
\includegraphics[height=0.43\columnwidth]{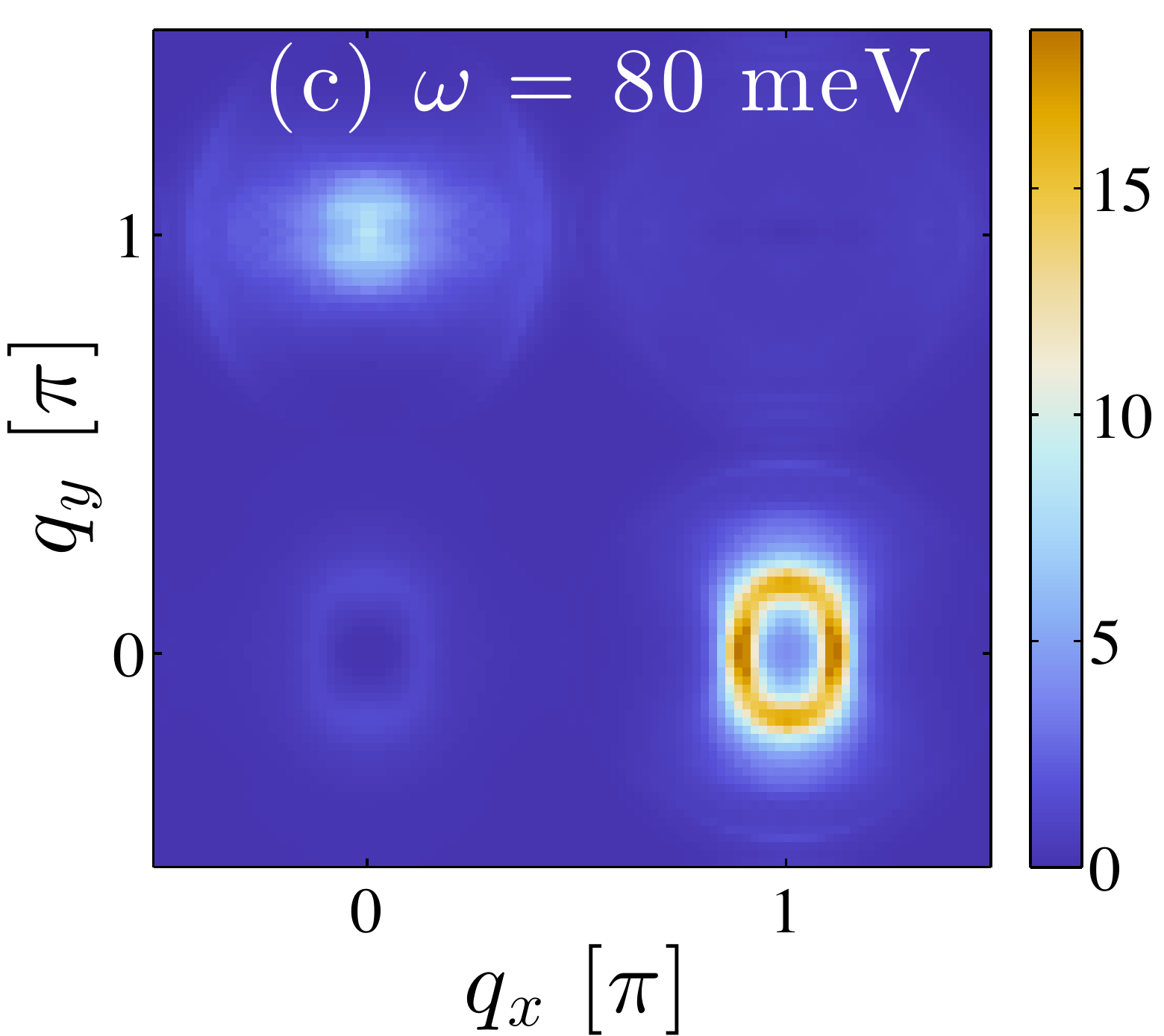}
\includegraphics[height=0.43\columnwidth]{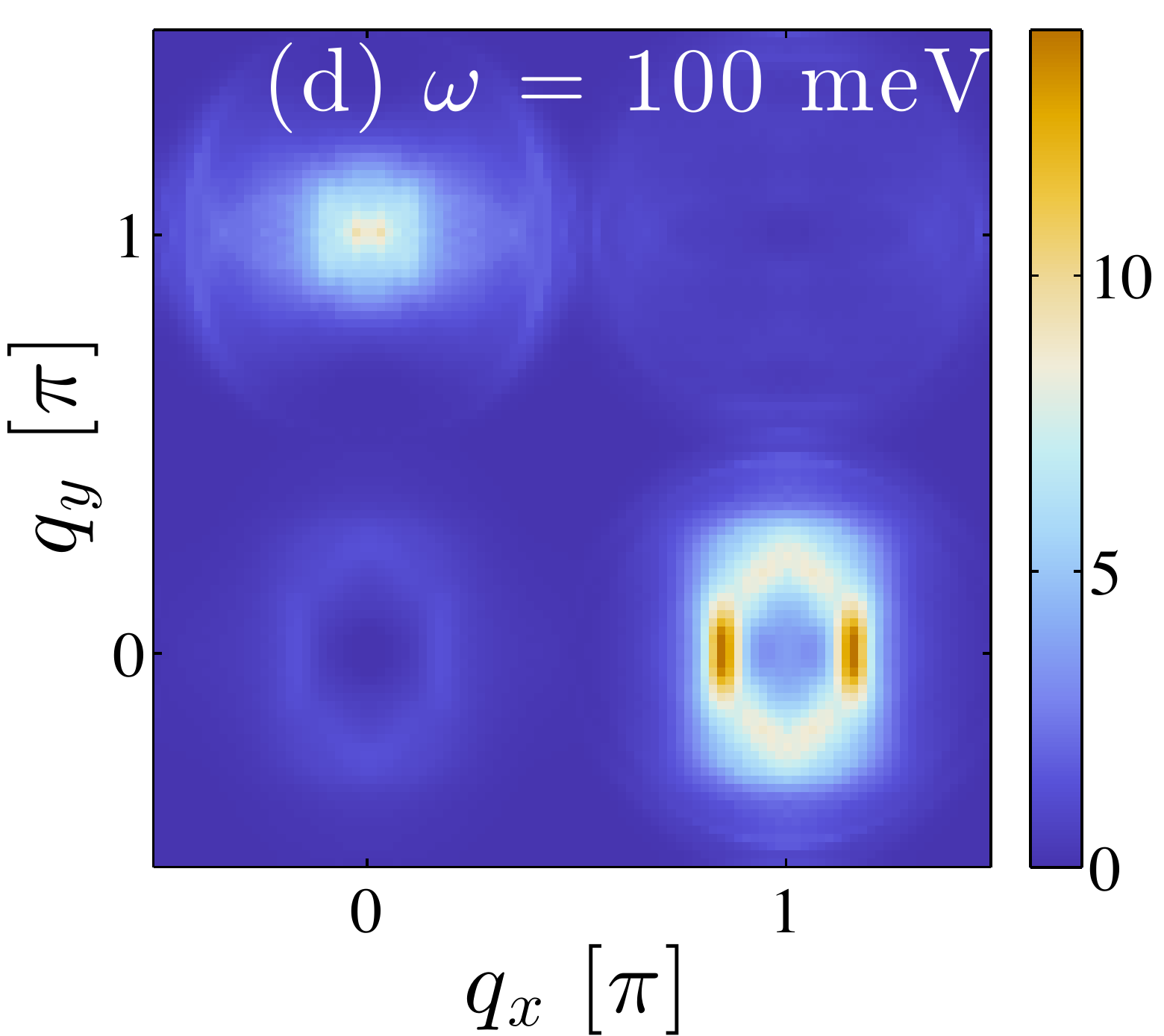}
\includegraphics[height=0.43\columnwidth]{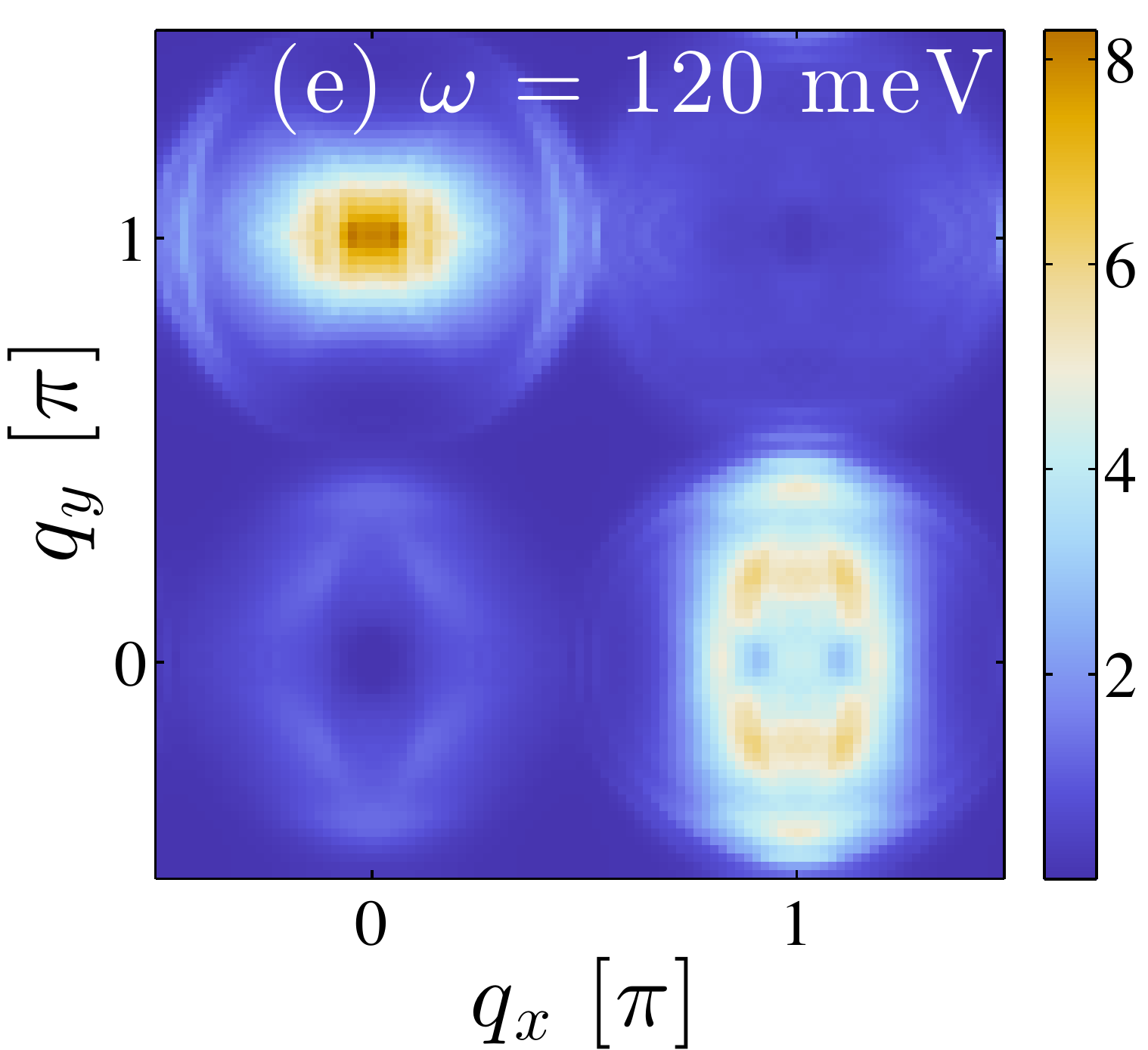}
\includegraphics[height=0.43\columnwidth]{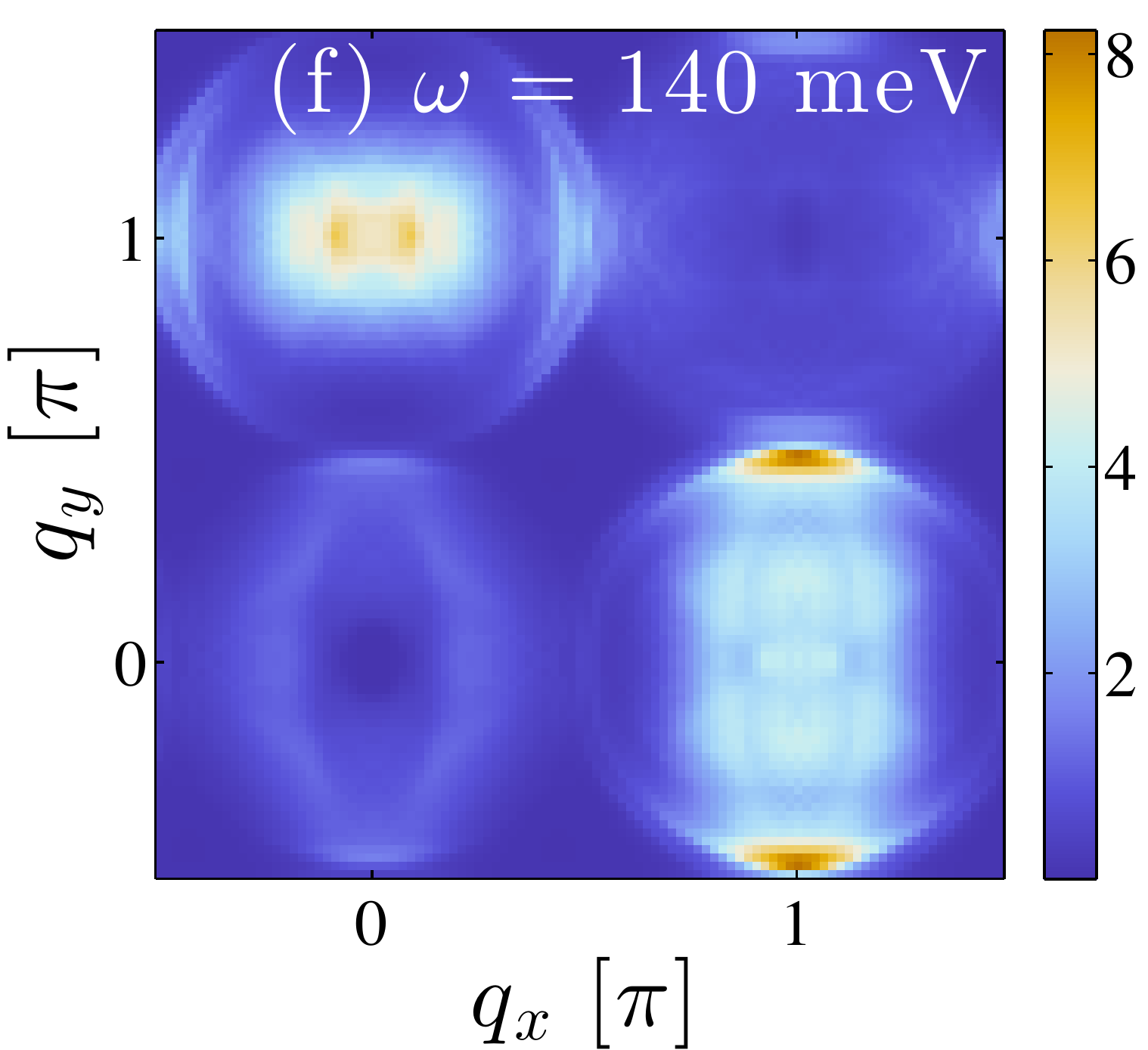}
 \caption{(Color online) Constant energy $\omega$ cuts of $\mbox{Im} \chi^{+-}_{\text{RPA}}({\mathbf q},\omega)$ in the SDW phase with $U = 1.1$ eV and $J = U/6$ shown at $\omega=20$ meV (a), $\omega=40$ meV (b), $\omega=80$ meV (c), $\omega=100$ meV (d), $\omega=120$ meV (e), and $\omega=140$ meV (f).}
 \label{fig:SDWU112Dcuts}
\end{figure}

The gapping of the particle-hole continuum at momenta near, for example, the $X$ point to higher energies with increasing $U$ can be seen explicitly from Fig.~\ref{fig:dispvsU0}, which shows the imaginary part of the bare physical susceptibility $\chi_0^{+-}({\mathbf q},\omega)$ for the same parameters as in Fig.~\ref{fig:dispvsU}. The effect of the particle-hole continuum has been analyzed previously by several other theoretical studies,\cite{kaneshita10} including the exitonic scenario,\cite{brydon09,knolle10,knolle11} and also studied in detail within the simpler one-band case.\cite{brydon09,schrieffer89,chubukov92,wenya12} For all the cases of $U$ used in Fig.~\ref{fig:dispvsU}, the system is metallic and contains a Fermi surface. This is seen explicitly from Fig.~\ref{fig:FS} where we show the Fermi surfaces  in the RBZ for all the cases corresponding to Fig.~\ref{fig:dispvsU} and Fig.~\ref{fig:dispvsU0}. As expected, the SDW order reconstructs the original pockets and generates small Dirac-like pockets at intermediate values of $U$ as discussed previously.\cite{ran09,luo10} Note, that the present band contains a hole pocket at $(\pi,\pi)$ in the normal (undoped) state which is the reason for the fact that the electron pocket at $Y$ also gets reconstructed. It is also evident from Fig.~\ref{fig:FS} that all values of $U$ applied in this study exhibit metallic bands with a Fermi surface. Therefore, $\mbox{Im}\chi^{+-}_0({\mathbf q},\omega)$ exhibits low-energy spectral weight at selected momentum regions corresponding to particle-hole scattering between the remaining parts of the reconstructed Fermi surface.

\begin{figure}[t]
\includegraphics[height=0.42\columnwidth]{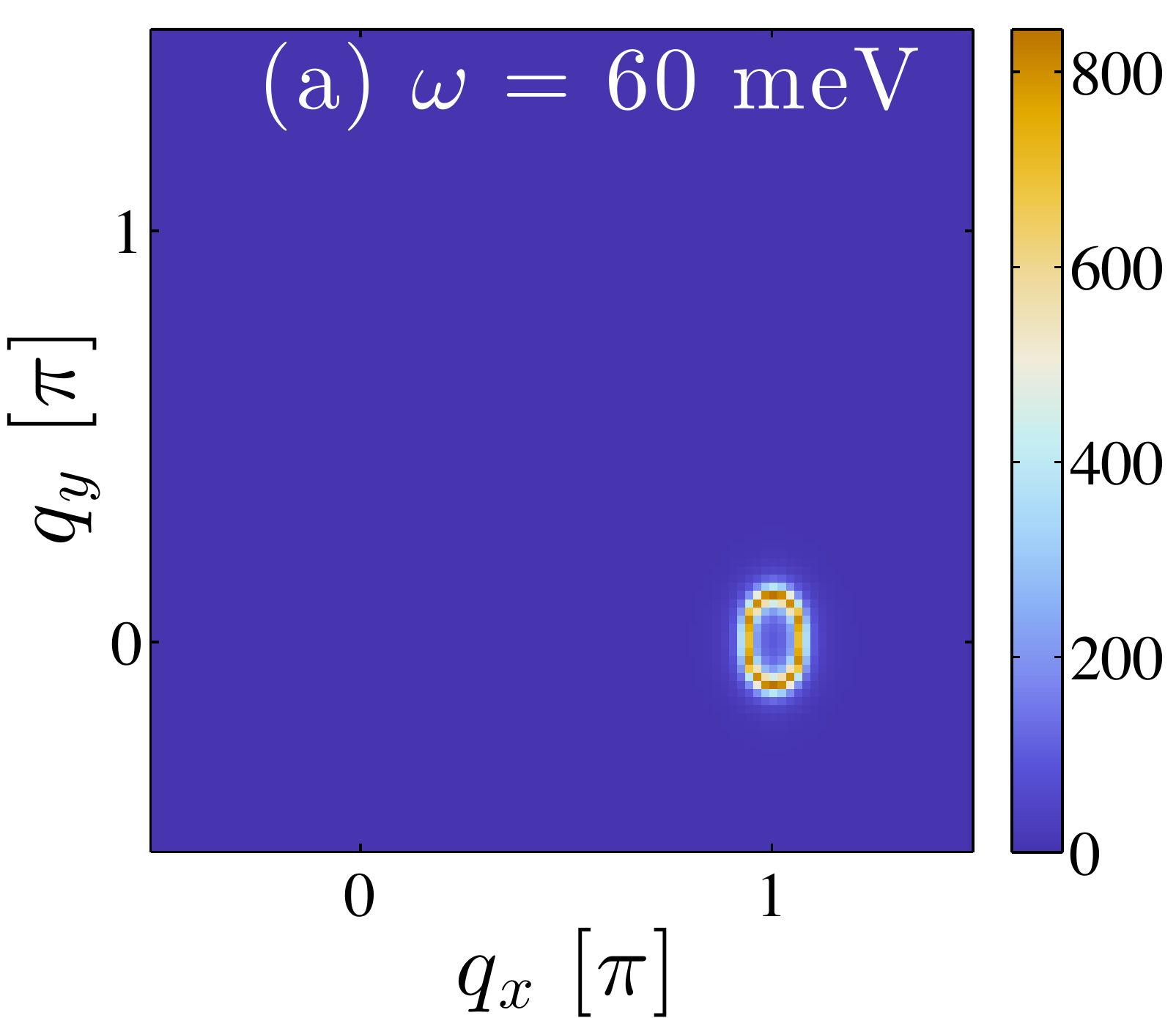}
\includegraphics[height=0.42\columnwidth]{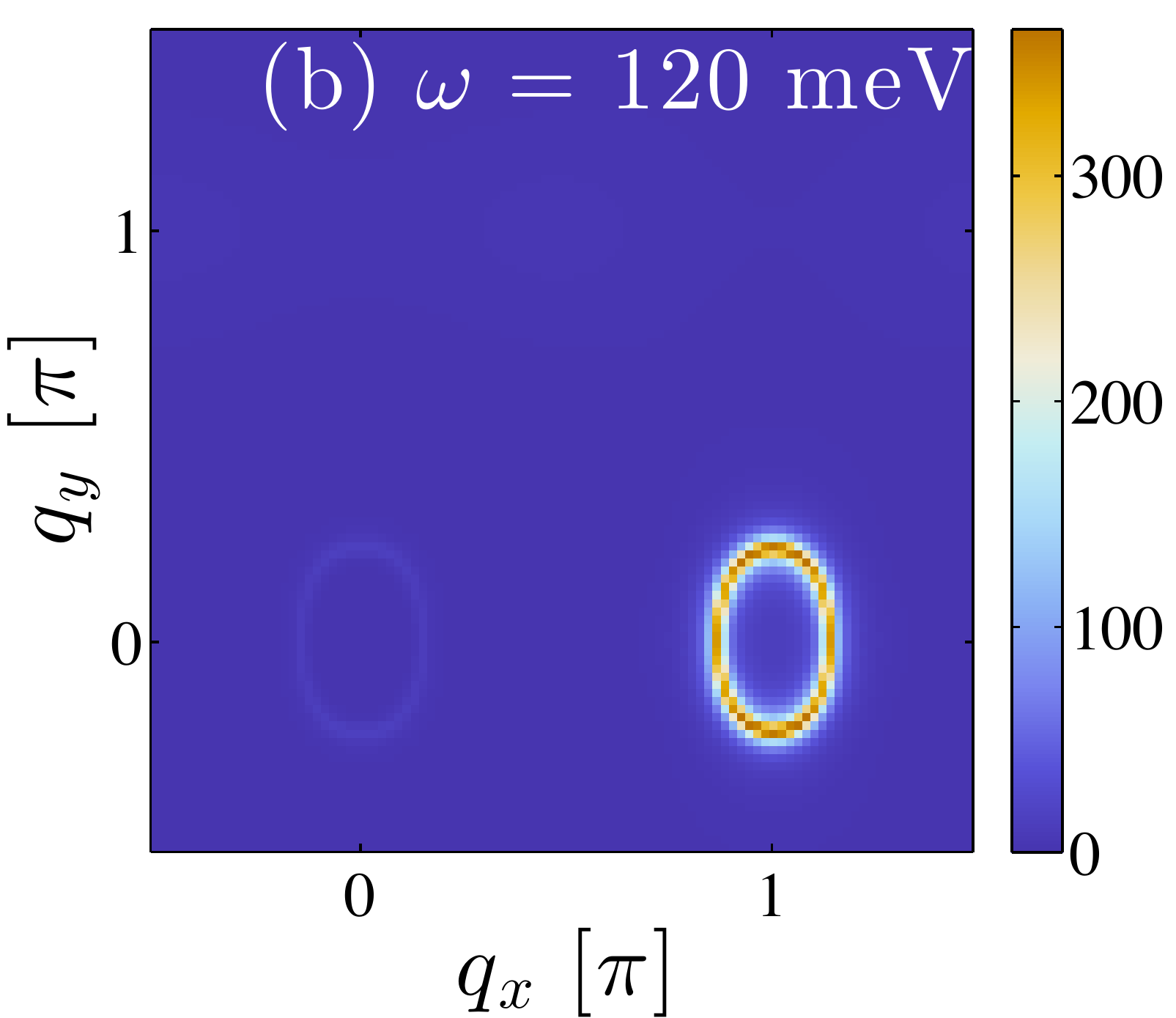}
\includegraphics[height=0.42\columnwidth]{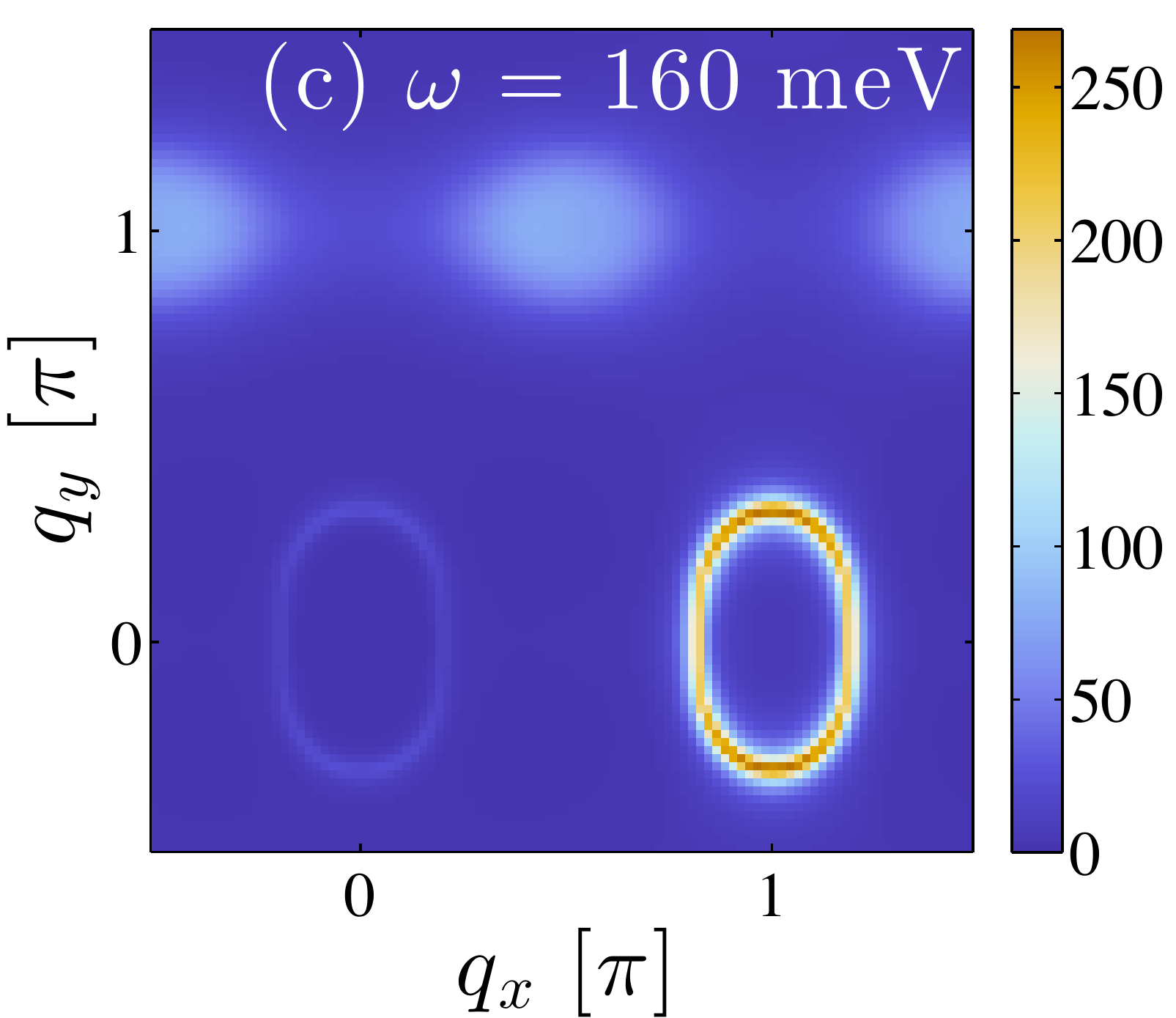}
\includegraphics[height=0.42\columnwidth]{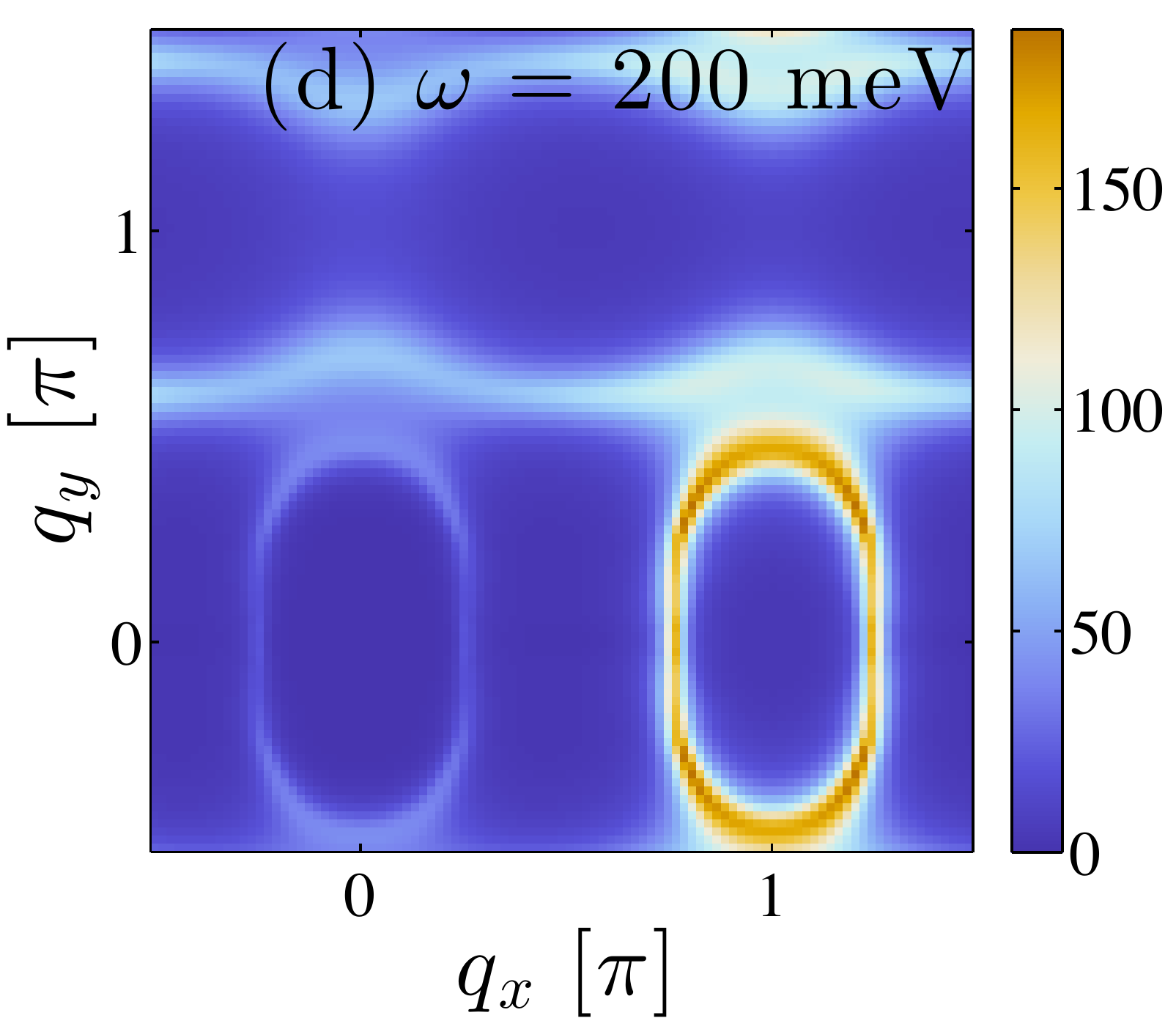}
\includegraphics[height=0.42\columnwidth]{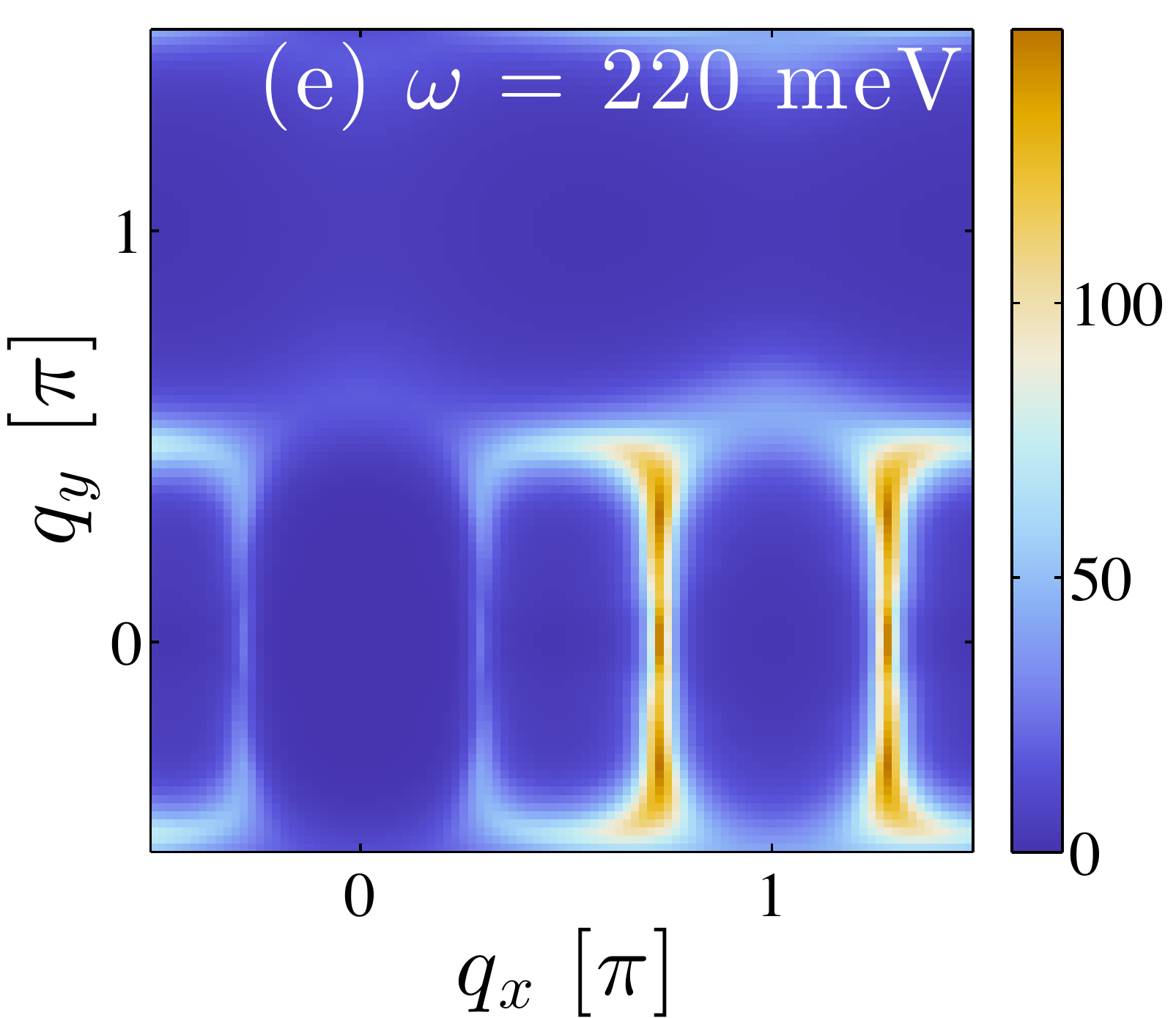}
\includegraphics[height=0.42\columnwidth]{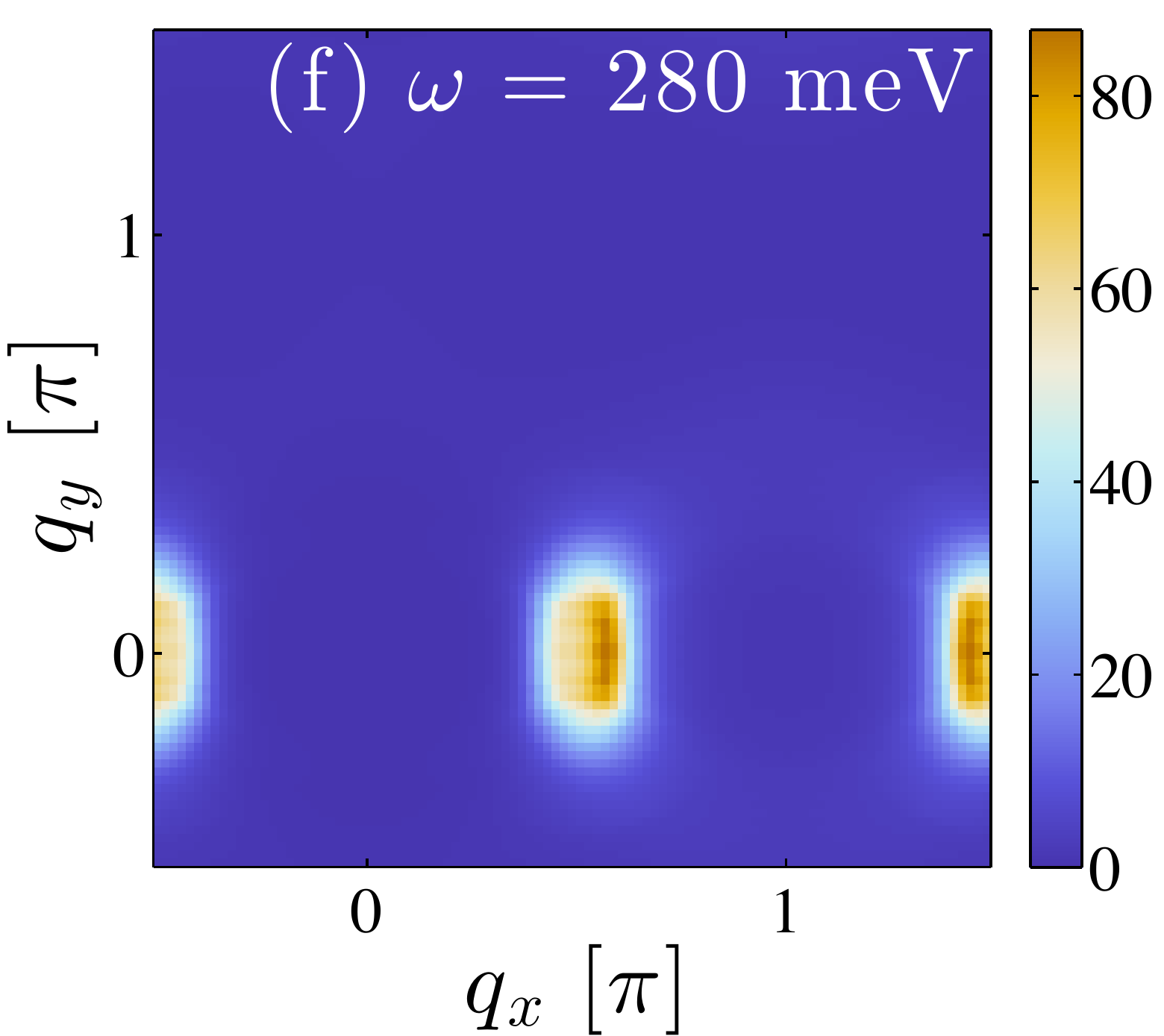}
 \caption{(Color online) Constant energy $\omega$ cuts of $\mbox{Im} \chi^{+-}_{\text{RPA}}({\mathbf q},\omega)$  in the SDW phase with $U=2.0$ eV and $J=U/6$ shown at $\omega=60$ meV (a), $\omega=120$ meV (b), $\omega=160$ meV (c), $\omega=200$ meV (d), $\omega=220$ meV (e), and $\omega=280$ meV (f).}
 \label{fig:SDWU202Dcuts}
\end{figure}

Focussing on the momentum structure of the spin susceptibility, we show in Fig.~\ref{fig:SDWU112Dcuts} and Fig.~\ref{fig:SDWU202Dcuts} the weight of the spin modes at fixed energy in momentum space for $U=1.1$ eV and $U=2.0$ eV, respectively. In the low-energy regime, both cases display an elliptic ring of spin waves caused by the anisotropic spin wave velocities in the SDW phase. The anisotropy of the spin response at $\mathbf{Q}_1$ is present already in the bare susceptibility and is directly related to the ellipticity of the electron pockets. At higher energies, there is a clear difference between the fate of the spin waves in the low-$U$ versus large-$U$ limits. In the latter case, the main weight is eventually exhibited along the antiferromagnetic $q_x$ direction as seen from panels (e) and (f) in Fig.~\ref{fig:SDWU202Dcuts}, very similar to the behavior of the spin dynamics obtained from the isotropic Heisenberg model.\cite{harriger11} By contrast, in the low-$U$ case the high-energy part of the spin modes is opposite in the sense that the damping eventually mainly affects the $q_x$-dispersing part of the modes leading to the spots along the ferromagnetic direction displaced from the $\mathbf{Q}_1$ point, as shown in Fig.~\ref{fig:SDWU112Dcuts}(e,f). Note, however, the existence of an intermediate energy regime where the largest weight is found along the antiferromagnetic $q_x$ direction as seen from Fig.~\ref{fig:SDWU112Dcuts}(d).

\begin{figure}[t]
\includegraphics[width=0.99\columnwidth]{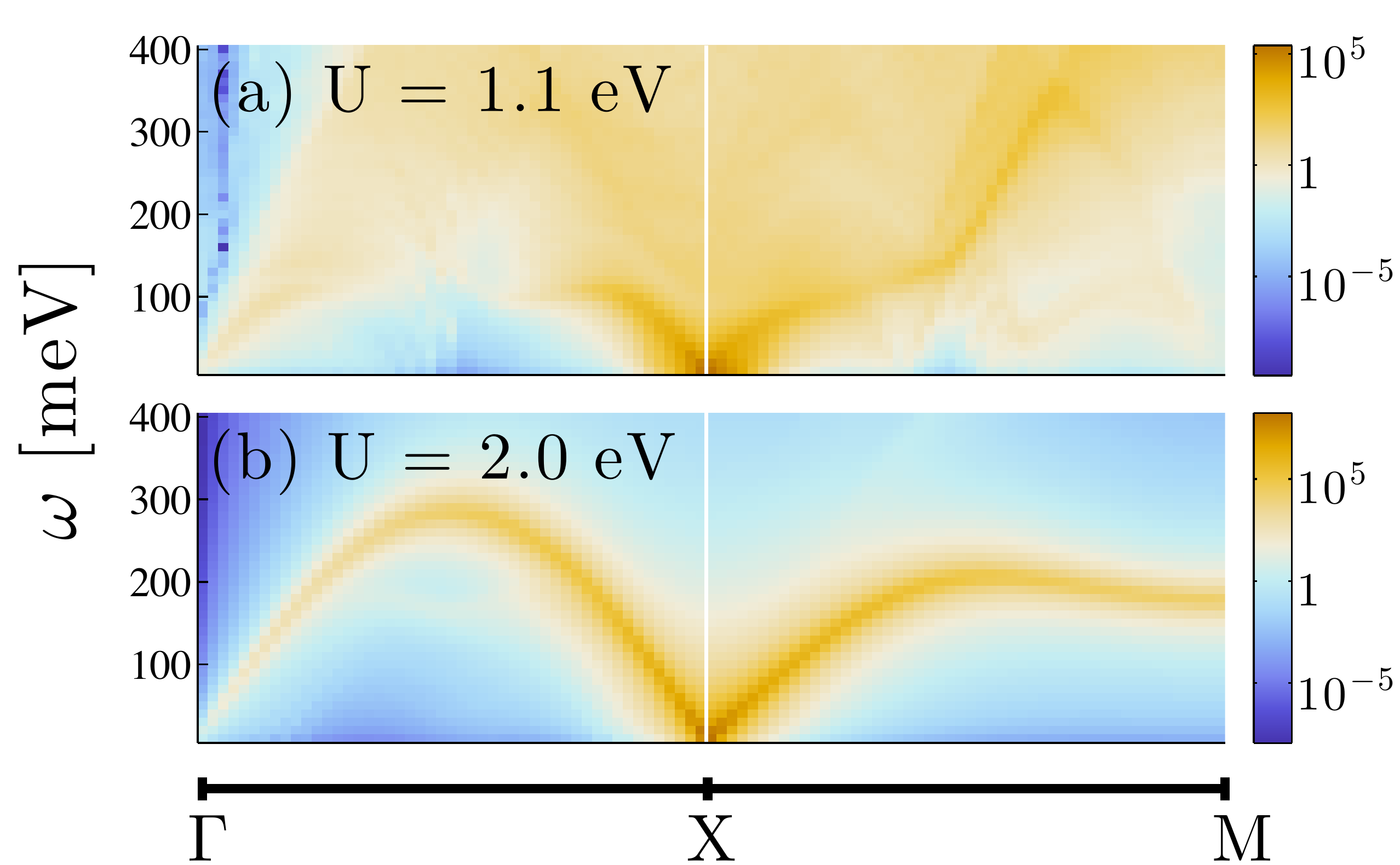}
\includegraphics[width=0.99\columnwidth]{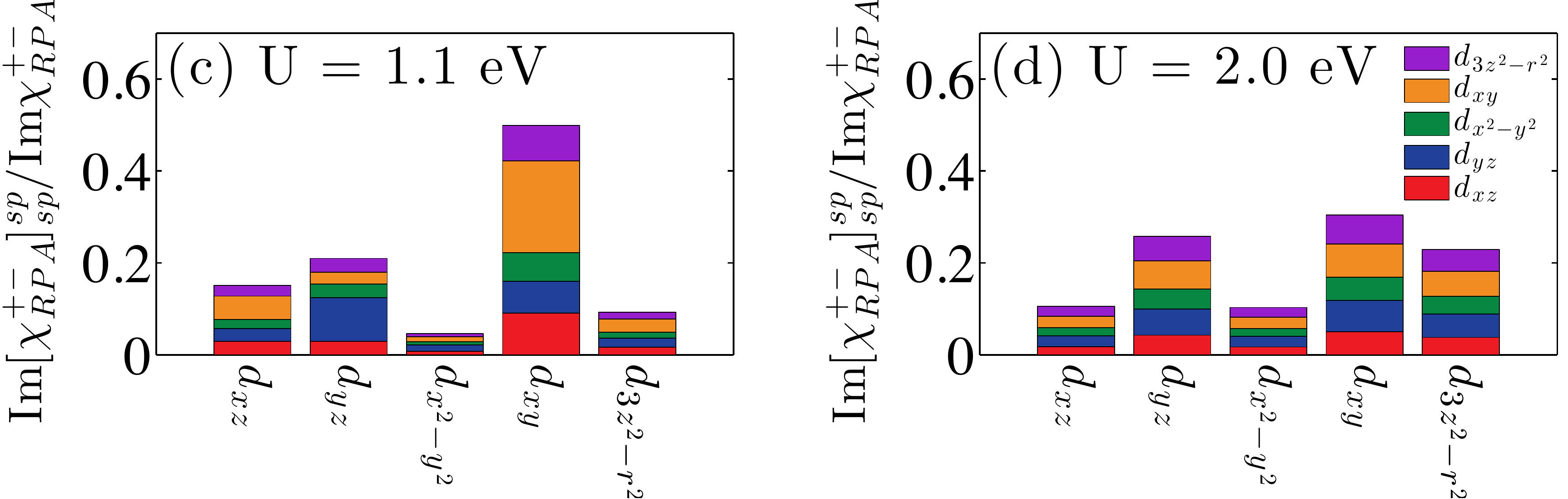}
 \caption{(Color online) Imaginary part of the physical transverse dynamical RPA susceptibility, $\mbox{Im} \chi^{+-}_{\text{RPA}}({\mathbf q},\omega)$, in the SDW phase along the perpendicular momentum cut $\Gamma \rightarrow X \rightarrow M$ for (a) $U=1.1$ eV and (b) $U=2.0$ eV on a log scale. Panels (c) and (d) reveal the spectral weight of the physical matrix components $\mbox{Im}[\chi_{RPA}^{+-}]^{sp}_{sp}$ (orbital $s$ is represented by color and $p$ can be read off the $x$-axis) for $U=1.1$ eV (c) and $U=2.0$ eV (d) contributing to the entire spectrum shown in (a) and (b), respectively.}
 \label{fig:perpcuts}
\end{figure}

In Fig.~\ref{fig:perpcuts}(a)-(b) we compare in greater detail the anisotropic dispersion of the spin excitations for both $U=1.1$ eV and $U=2.0$ eV by displaying $\mbox{Im} \chi^{+-}_{\text{RPA}}({\mathbf q},\omega)$ along the perpendicular momentum path $\Gamma \rightarrow X \rightarrow M$ on a log scale, where $M=(\pi,\pi)$. As seen, the $q_x-q_y$ anisotropy of the spin excitations around $X$ is clearly seen, especially in the case of $U=2.0$ eV. Furthermore, one sees that for $U=1.1$ eV, the broadening is larger along $X \rightarrow M$ in agreement with Fig.~\ref{fig:SDWU112Dcuts}(c,d), whereas the detailed final structure of the spectral weight shown in Fig.~\ref{fig:SDWU112Dcuts}(e,f) is a result of an interplay between overlapping spin modes and a strong overlap with the particle-hole continuum.

The orbitally resolved weight of the spin waves can be easily extracted within the formalism presented here, and is of relevance to resonant inelastic x-ray scattering (RIXS).\cite{ament11}  Figure~\ref{fig:perpcuts}(c,d) display the intra- and inter-orbitally resolved total momentum ${\bf q}$ and energy $\omega$ summed spectral weight in the $({\bf q},\omega)$-range shown in Fig.~\ref{fig:perpcuts}(a,b), respectively. We find that in the limit of weak $U$, the broad spin waves are almost entirely of $d_{xy}$ character, with the dominant contribution coming from the intra-orbital component as seen in Fig.~\ref{fig:perpcuts}(c). By contrast, for large $U$ we find that none of the physical orbitally resolved intra- or inter-orbital susceptibilities are negligible, but most of the spectral weight in the spin waves around $X$ comes from scattering processes involving the $d_{yz}$, $d_{xy}$, and $d_{3z^2-r^2}$ orbitals as seen from Fig.~\ref{fig:perpcuts}(d). In addition, it can be seen also from Fig.~\ref{fig:perpcuts}(d) that in the case of large $U$, the majority of the weight originates from inter-orbital susceptibilities. Both results shown in Fig.~\ref{fig:perpcuts}(c,d) are qualitatively different from the orbital content of the spin waves inferred from an earlier  two-orbital study.\cite{knolle11}

The results presented in Figs.~\ref{fig:dispvsU}-\ref{fig:SDWU202Dcuts} clearly show the parameter-sensitivity of the high-energy modes and their associated damping. We expect a similar dependence of the details on the spectral weight to the original band structure applied in $\mathcal{H}_0$ since that obviously influences the bounds and strengths of the particle-hole continuum, but we have not further explored the band dependence of the high energy spin response here. From this perspective, however, it seems consistent that even among the 122 materials there is significant differences between the behavior of the high-energy susceptibility, with BaFe$_2$As$_2$ exhibiting a larger directional dependent damping\cite{harriger11} compared to, for example CaFe$_2$As$_2$ where ring-like spin wave dispersions are observed to high energies well above $\sim 100$ meV.\cite{diallo09,zhao09}

\subsection{Spin excitations in the nematic phase}

We turn now to a study of the paramagnetic phase motivated largely by recent neutron experiments in the uniaxial-strained tetragonal phase of BaFe$_{2-x}$Ni$_x$As$_2$.\cite{lu14} There it was found that uniaxial strain causes a large change of the low-energy spin fluctuations from $C_4$ to $C_2$ symmetric. More specifically, the scattering intensity at $\mathbf{Q}_1$ at 6 meV was observed to increase significantly upon approaching the magnetic transition temperature $T_N$ from above, whereas the corresponding intensity at $\mathbf{Q}_2$ remained largely unchanged (see e.g. Figs. 3 and 4 of Ref. \onlinecite{lu14}). 

\begin{figure}[t]
\includegraphics[height=0.4\columnwidth]{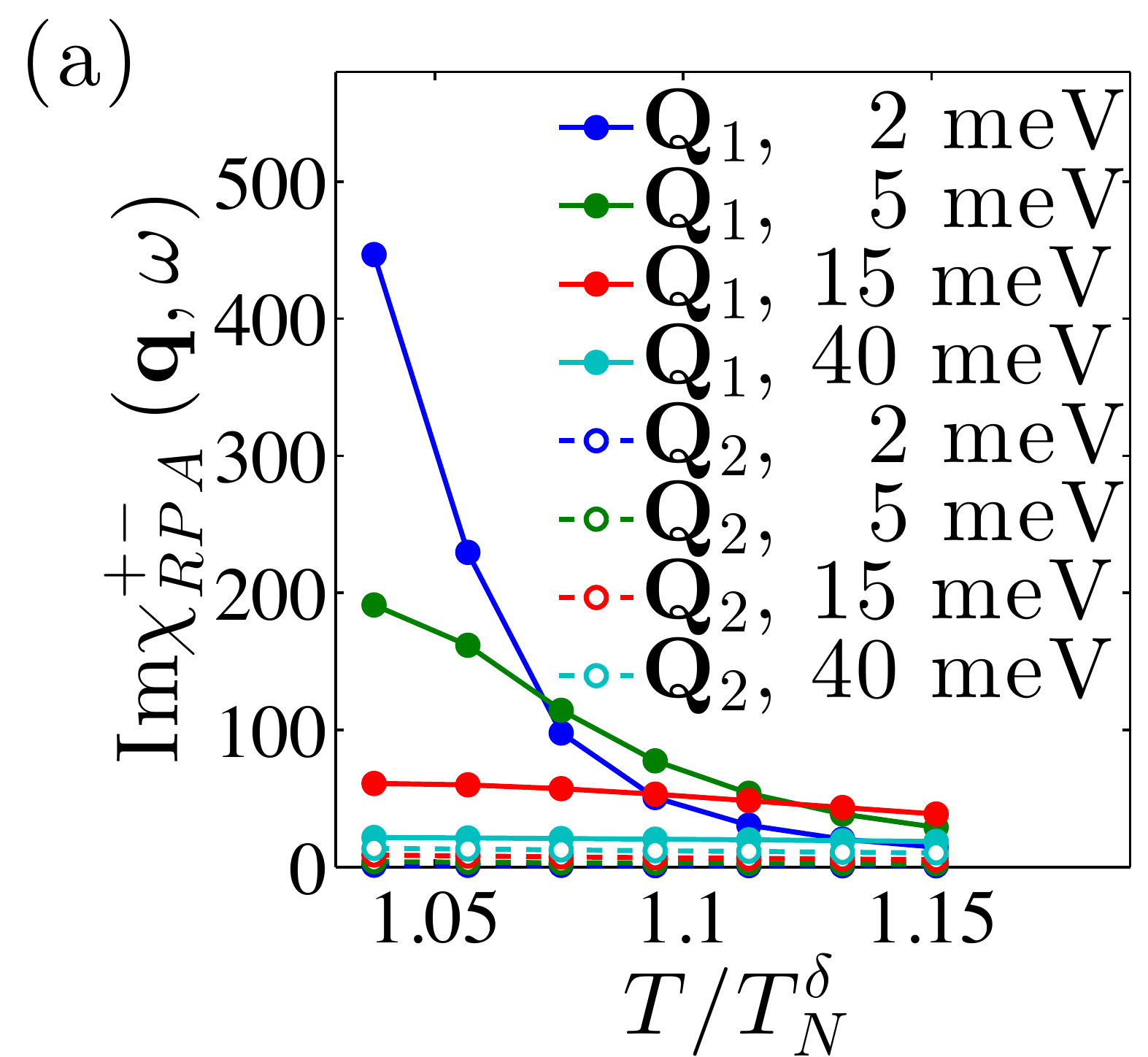}
\includegraphics[height=0.4\columnwidth]{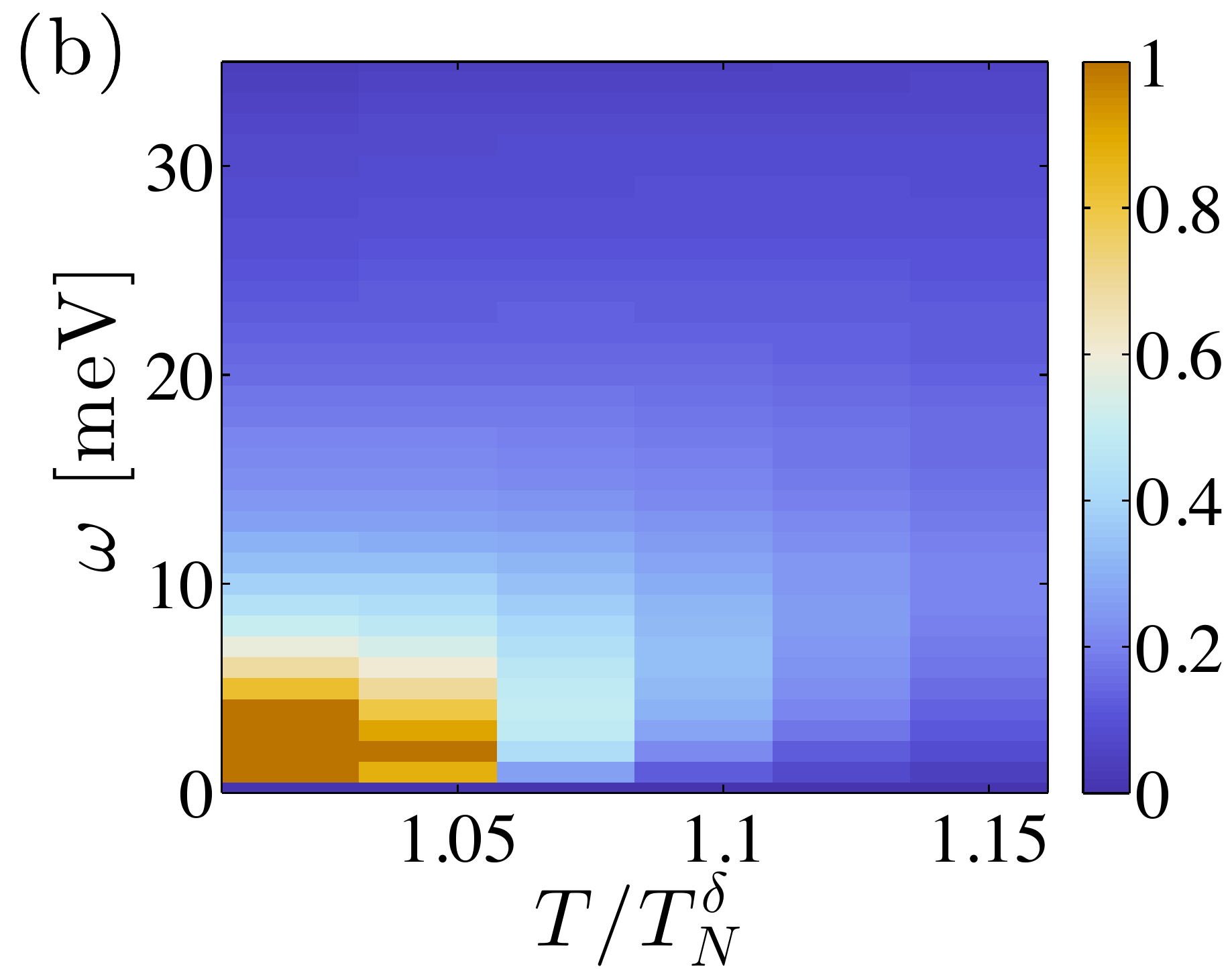}
 \caption{(Color online) (a) Spin susceptibility $\mbox{Im} \chi^{+-}_{\text{RPA}}({\mathbf q},\omega)$ in the paramagnetic nematic phase at $\mathbf{Q}_1=(\pi,0)$ and $\mathbf{Q}_2=(0,\pi)$ versus temperature $T$ for representative energies shown in the legend and an orbital splitting of $\delta=65$ meV. (b) 2D map of the normalized spin anisotropy, $(\mbox{Im} \chi_{\text{RPA}}(\mathbf{Q}_1,\omega)-\mbox{Im} \chi_{\text{RPA}}(\mathbf{Q}_2,\omega))/\cal{N}$ where ${\cal{N}}=\mbox{Max} [\mbox{Im} \chi_{\text{RPA}}(\mathbf{Q}_1,\omega) - \mbox{Im} \chi_{\text{RPA}}(\mathbf{Q}_2,\omega)]$ at $T=1.05T_N^\delta$.}
 \label{fig:nematic}
\end{figure}

\begin{figure}[hb!]
\includegraphics[height=0.43\columnwidth]{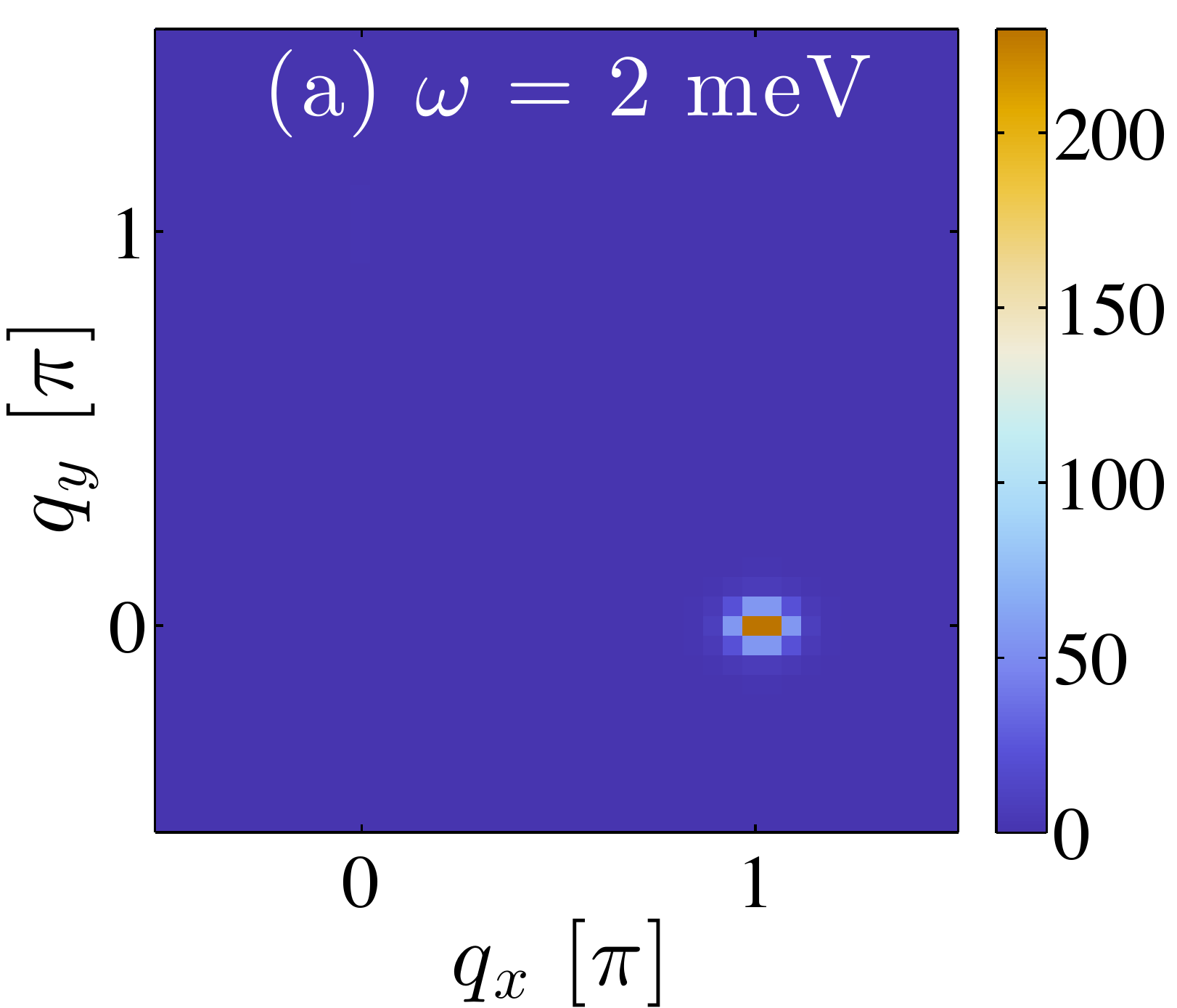}
\includegraphics[height=0.43\columnwidth]{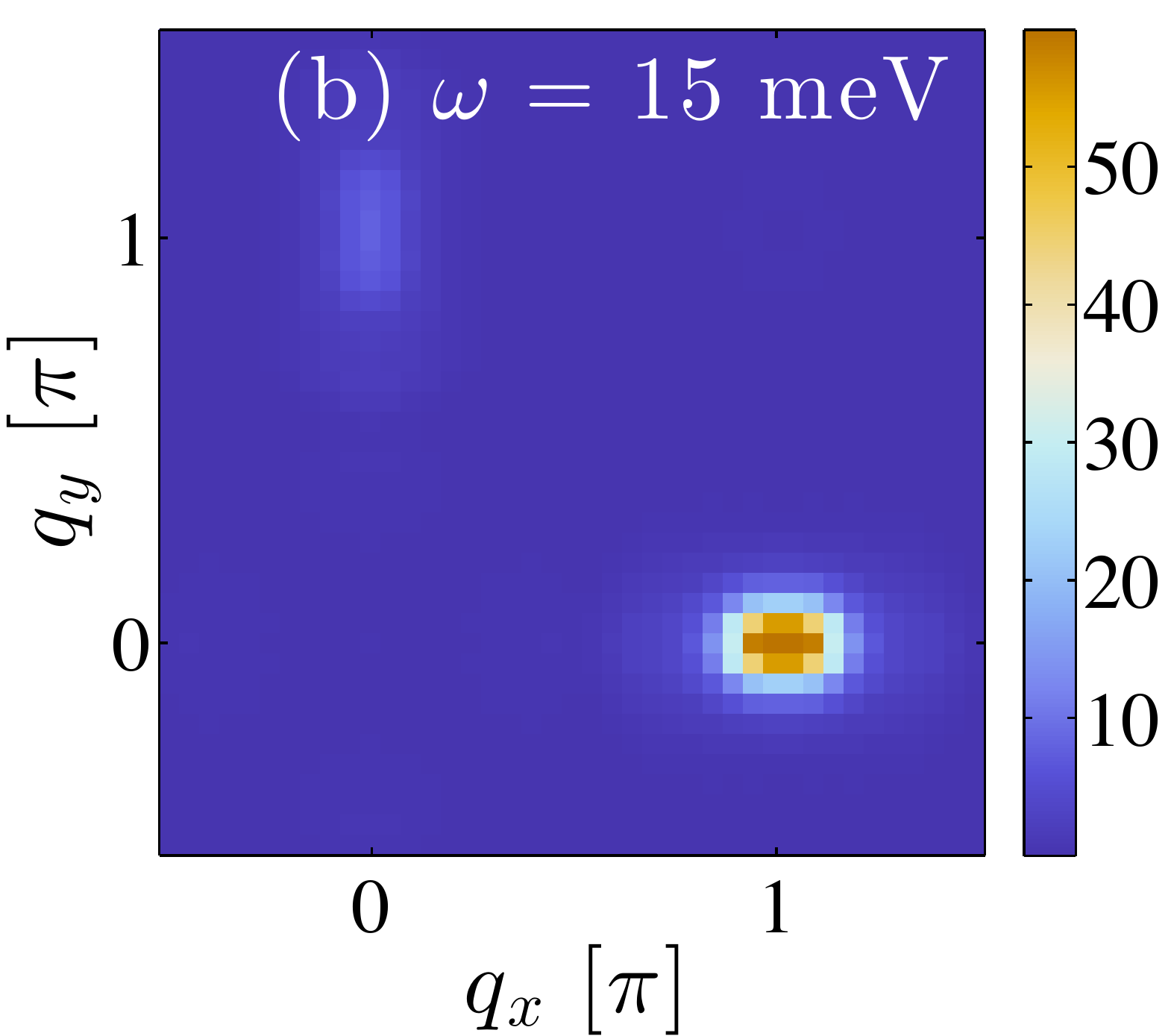}
\includegraphics[height=0.43\columnwidth]{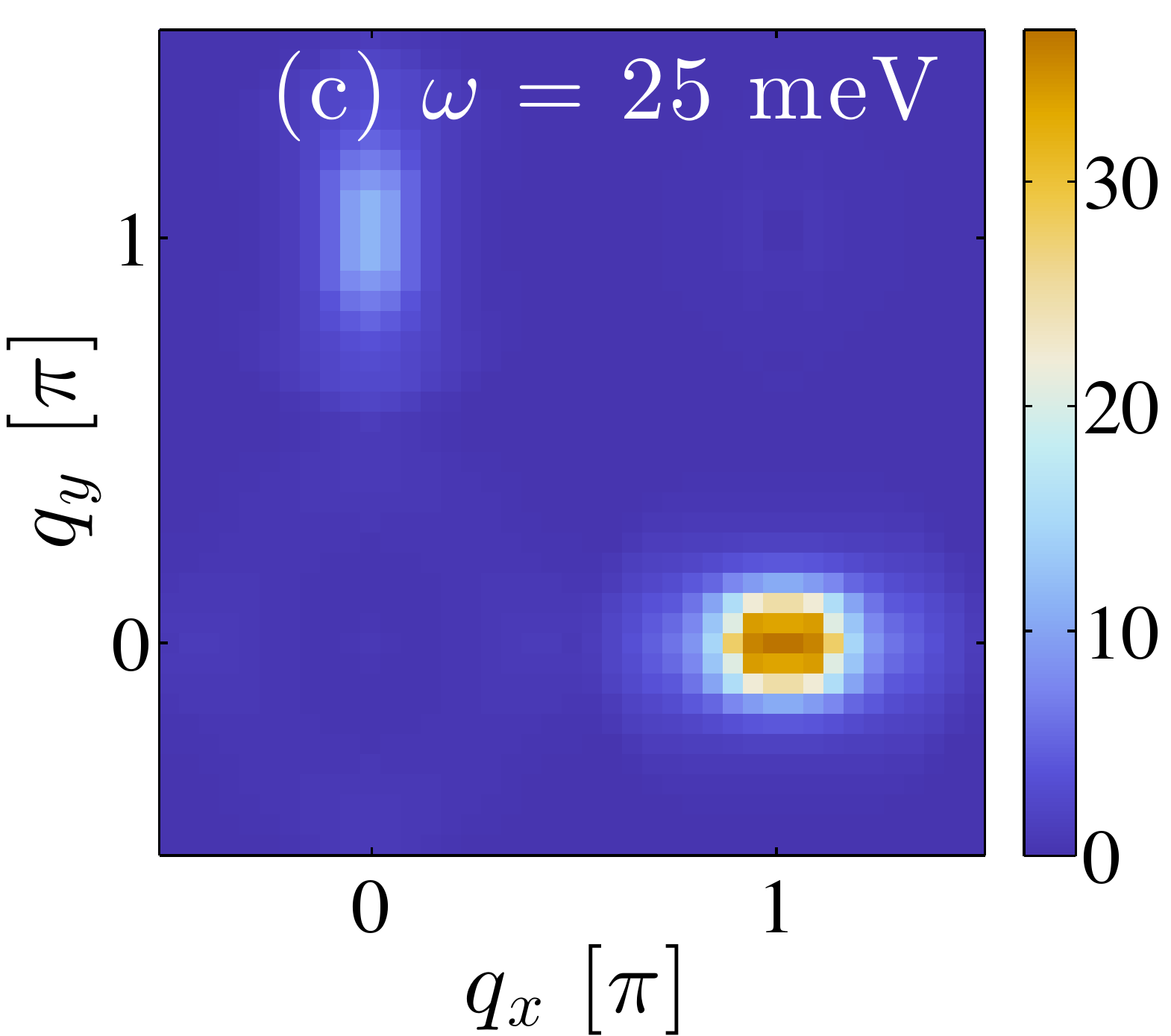}
\includegraphics[height=0.43\columnwidth]{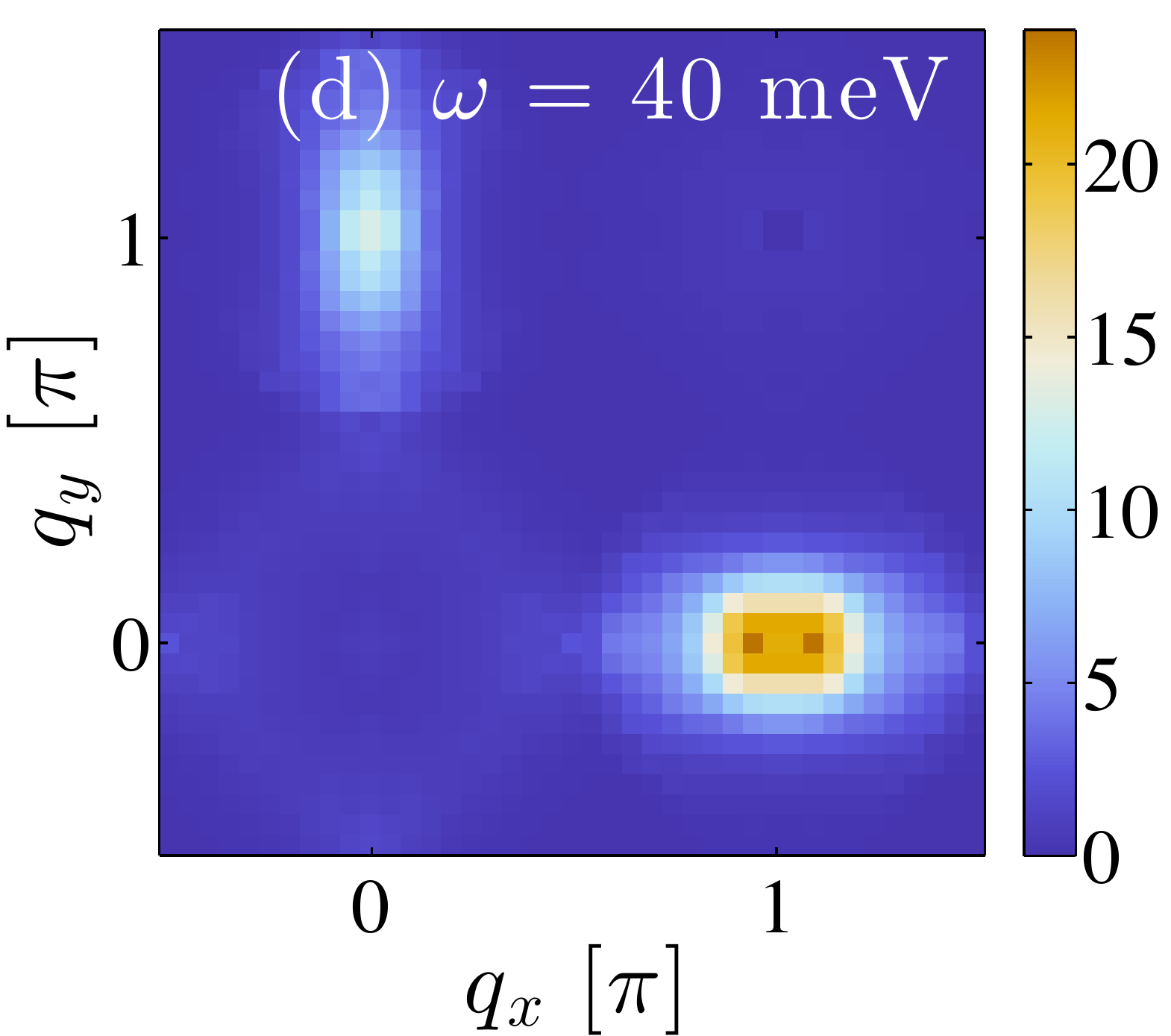}
\includegraphics[height=0.43\columnwidth]{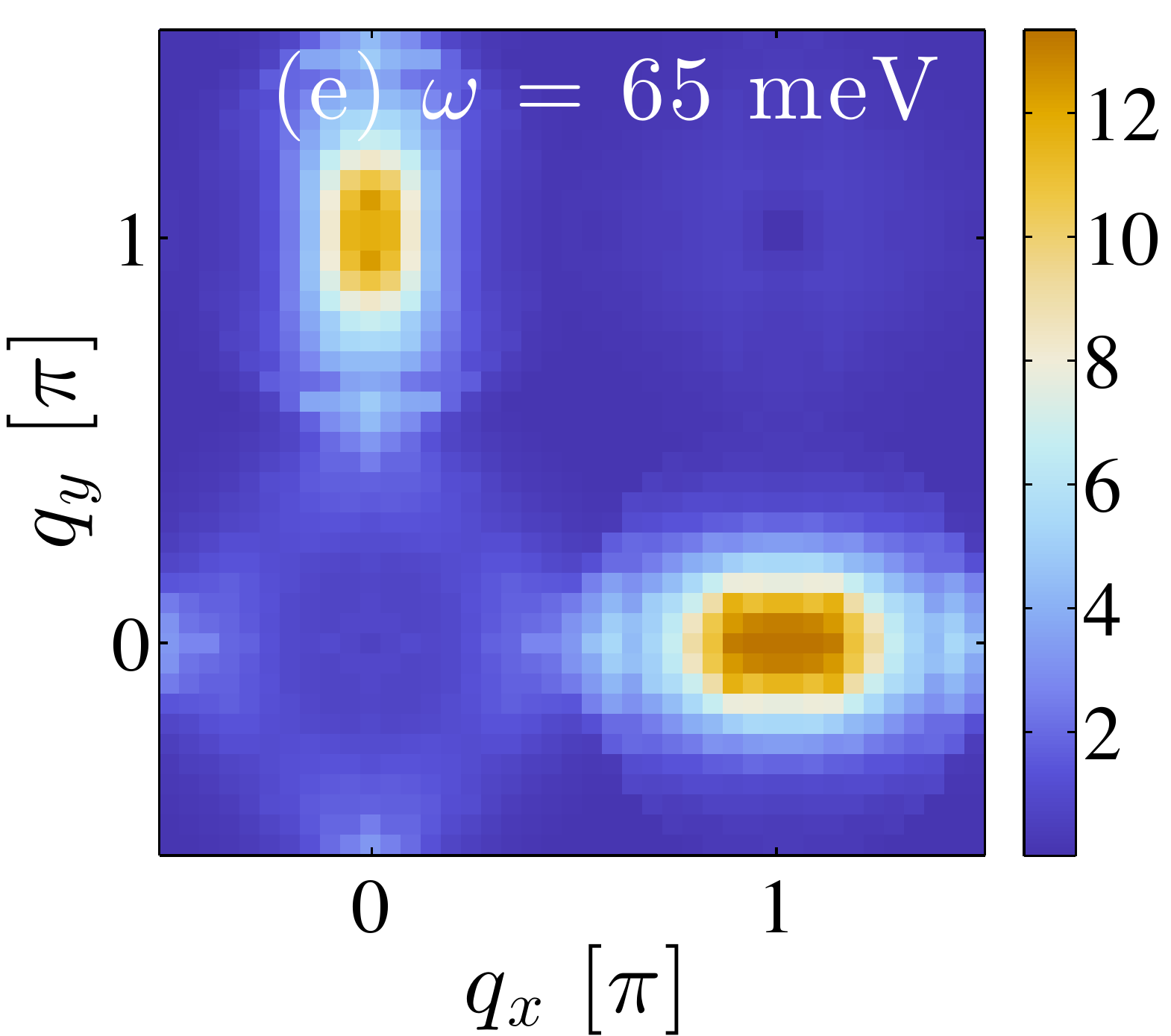}
\includegraphics[height=0.43\columnwidth]{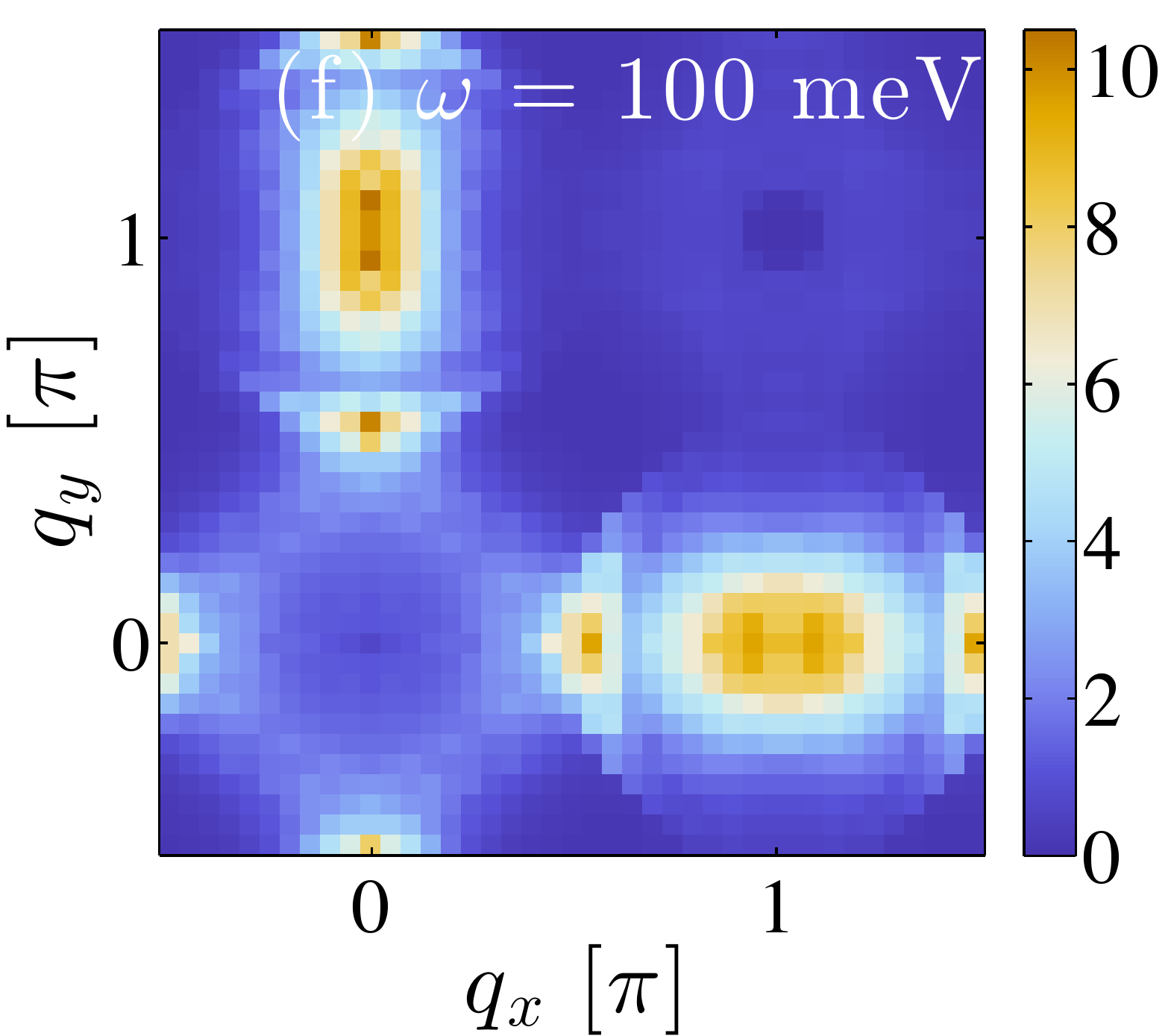}
\caption{(Color online) Constant energy cuts of the spin susceptibility $\mbox{Im} \chi^{+-}_{\text{RPA}}({\mathbf q},\omega)$ in the paramagnetic nematic phase at temperature $T=1.04T_N^\delta$ for $U=1.0$ eV, $J=U/4$, and $\delta = 65$ meV shown at $\omega=2$ meV (a), $\omega=15$ meV (b), $\omega=25$ meV (c), $\omega=40$ meV (d), $\omega=65$ meV (e), and $\omega=100$ meV (f).}
 \label{fig:nematic2D}
\end{figure}

To mimic the nematic phase we impose $m_{\mu\nu} = 0$ ($\hat{M} = \mathbf{0}$) and introduce an explicit orbital splitting given by
\eq{
	\mathcal{H}_{oo} = \frac{\delta}{2} \sum_{k}\Big( n_{yz}(\k) - n_{xz}(\k)\Big) \ ,
}{eq:Hoo}
motivated by the observed energy splitting $\delta$ between the $d_{xz}$ and $d_{yz}$ orbitals by ARPES.\cite{yi11} We stress that the approach here is simply to investigate the consequences of an orbital splitting on the spin excitations, and not the origin of the orbital splitting itself. In Fig.~\ref{fig:nematic}(a) we show the $T$-dependence of the imaginary part of $\chi^{+-}_{\text{RPA}}({\mathbf q},\omega)$ in the paramagnetic phase for both $\mathbf{Q}_1$ and $\mathbf{Q}_2$ at different energies $\omega$. The orbital splitting slightly enhances the N\'{e}el transition temperature compared to the case with $\delta=0$, and is therefore denoted by $T_N^\delta$. As seen, the anisotropy increases dramatically at low energies upon approaching $T_N^\delta$, similar to the findings in earlier studies relevant to the cuprates,\cite{andersen12} and also two recent studies of iron pnictides.\cite{gastiasoro14,su14} This property is mapped out in more detail in Fig.~\ref{fig:nematic}(b) which clearly shows the confined region of spin anisotropy in both $\omega$ and $T$. For the results presented in Fig.~\ref{fig:nematic} we have used an orbital splitting of $\delta=65$ meV which sets the size of the anisotropy region in Fig.~\ref{fig:nematic}(b); a smaller $\delta$ diminishes this region but the arbitrarily large spin anisotropy persists when approaching $T_N^\delta$ which is simply caused by the non-linear divergence of the RPA denominator, picking out $\mathbf{Q}_1$ as the dominant instability due to the chosen sign of $\delta$. From Fig.~\ref{fig:nematic}(b) it is evident that in twinned samples the total spin response, $\mbox{Im} \chi_{\text{RPA}}(\mathbf{Q}_1,\omega)+\mbox{Im} \chi_{\text{RPA}}(\mathbf{Q}_2,\omega)$, will also display a strong upturn at low $\omega$ upon approaching $T_N^\delta$.\cite{zhang14}

In Fig.~\ref{fig:nematic2D} we show the momentum dependent constant energy cuts for the same parameters as in Fig.~\ref{fig:nematic} at temperature $T=1.04T_N^\delta$. As seen, the very prominent spin anisotropy at low energies gradually vanishes as the energy is enhanced and the spin excitations become increasingly broad around both $\mathbf{Q}_1$ and $\mathbf{Q}_2$. Future neutron scattering measurements should be able to measure the vanishing of the spin anisotropy as a function of energy, and its quantitative behavior may give important clues to the origin of the nematic electronic behavior of the paramagnetic phase.

Recently a direct link between transport anisotropy and spin excitation anisotropy was demonstrated by Lu {\it et al.},\cite{lu14} finding the same onset temperature for anisotropy of these two distinct properties. An explanation for this has been given in terms of strongly renormalized elastic scattering centers where pinning of the anisotropic spin fluctuations leads to emergent highly anisotropic impurity centers as $T$ approaches $T_N$.\cite{gastiasoro14,yanwang14} These emergent defect scattering sites exhibit a pronounced elongated shape which directly causes a larger scattering rate for transport along the ferromagnetic $b$ axis compared to the antiferromagnetic $a$ axis in agreement with the transport experiments.\cite{tanatar10,chu10,ying11,chu12,blomberg13,ishida13,kuo14}

\section{conclusions}

We have studied the dynamical spin susceptibility as a function of Hubbard-Hund interactions within a five-orbital model relevant to iron pnictides. This allowed us to investigate the evolution of the spin response in the SDW phase from broad overdamped modes at low $U$ to sharply dispersing spin waves at higher $U$ but still in the metallic state. Both the structure of the spectral weight at high energies (in particular its anisotropy) and the orbital composition of the spin waves were shown to depend qualitatively on the strength of interactions. In the paramagnetic nematic phase modelled by an orbital splitting between the $d_{xz}$ and $d_{yz}$ orbitals, a strong spin anisotropy is present near the transition to the SDW phase at low energies, in agreement with recent neutron scattering measurements.\cite{lu14}

\section{Acknowledgements}

We thank E. Bascones, P. Dai, P. J. Hirschfeld, A. Kreisel, and S. Mukherjee for useful discussions.
We acknowledge support from a Lundbeckfond fellowship (grant A9318). 


\end{document}